\documentclass[amsmath,amssymb,notitlepage, twocolumn,aps,pra,floatfix]{revtex4-2}
\usepackage{xcolor}
\usepackage[colorlinks=true, allcolors = black]{hyperref}
\usepackage{url}
\usepackage{qcircuit}

\usepackage{subfigure}
\usepackage{graphicx}
\usepackage{dcolumn}
\usepackage{bm}

\usepackage{multirow}
\usepackage{braket}
\usepackage{amsmath}
\usepackage{cleveref}
\usepackage{mathtools}

\DeclareMathOperator{\Tr}{Tr}

\newcommand{\leftlabel}[1]{\push{\phantom{#1}}& \lstick{#1}}

\begin{document}
\preprint{APS/123-QED}

\title{Remote Cooling of Spin-ensembles through a Spin-mechanical Hybrid Interface}

\author{Yang Wang$^{1}$}
\author{Durga Bhaktavatsala Rao Dasari$^{1}$}\thanks{d.dasari@pi3.uni-stuttgart.de}
\author{J\"{o}rg Wrachtrup$^{1,2}$}

\affiliation{$^{1}$3. Physikalisches Institut, ZAQuant, University of Stuttgart, Allmandring 13, 70569 Stuttgart, Germany}
\affiliation{$^{2}$Max Planck Institute for Solid State Research, Heisenbergstraße 1, 70569 Stuttgart, Germany}

\begin{abstract}
\noindent We present a protocol for the ground-state cooling of a tripartite hybrid quantum system, in which a macroscopic oscillator acts as a mediator between a single probe spin and a remote spin ensemble. In the presence of weak dispersive coupling between the spins and the oscillator, cooling of the oscillator and the ensemble spins can be achieved by exploiting the feedback from frequent measurements of the single probe spin. We explore the parameter regimes necessary to cool the ensemble, the oscillator, or both to their thermal ground states. This novel cooling protocol shows that, even with only weak dispersive coupling, energy transfer-like effects can be obtained by simply manipulating the probe spin. These results not only contribute to the development of a practical solution for cooling/polarizing large spin ensembles, but also provide a relatively simple means of tuning the dynamics of a hybrid system. The proposed protocol thus has broader implications for advancing various quantum technology applications, such as macroscopic quantum state generation and remote sensing.
\end{abstract}
\maketitle

\section{Introduction}
 
\noindent Translating the potential of quantum technology into real-world applications remains a significant challenge due to the inherent limitations of current experimental platforms \cite{Battistel2023Challenges,Lordi2021}. One promising avenue is to explore the hybrid system approach, which promises to fully exploit the unique capabilities of different systems \cite{XiangRMP2013,Kurizki2015,HeiPRL2023,Kolkowitz2012Sensing}.  In particular, high-fidelity entanglement between different systems can be established by exploiting their common coupling to a transducer, thereby generating indirect interactions \cite{Rabl2010NPhysics, Kolkowitz2012Sensing}. 

In this work, we consider a tripartite hybrid system where an oscillator serves as a transducer for a single probe spin and a spin ensemble. By exploiting the measurement feedback from the frequently measured probe spin, efficient heat extraction from the ensemble can be realized, requiring only weak dispersive (off-resonant) coupling between the spins and the oscillator.   The probe control sequence is shown schematically in Fig.~\ref{fig:schematic_illustration}(b). Achieving complete ensemble polarization by such probe projections can avoid the dark state problem associated with resonant energy transfer. Furthermore, the ability to operate under dispersive interaction is important for various systems whose original dynamics do not allow direct energy exchange \cite{YangNatPhotonics2016,Browaeys2020}. Finally, we show that the same control sequence on the probe can also be used to prepare macroscopic oscillators (e.g., NMO) into specific quantum states that can potentially be used for sensing and error correction applications.

The oscillator here serves as a simplified conceptual model for a common bus (transducer) connecting two distant quantum systems that are not directly or weakly coupled. One can consider a general scenario where the bus is a large spin network, which can be effectively approximated as a bosonic system via the Holstein-Primakoff transformation, such as the spin wave excitations (magnons)  
\cite{ViolaKusminskiy2019,Bejarano2024}.  This scenario is particularly relevant when a single nitogen-vacancy (NV) electronic spin in diamond is used to polarize or probe distant target spins beyond the distance criteria set by the sensing volume \cite{Staudacher2013}. The target may be distant carbon nuclear spins within the diamond substrate \cite{Goldblatt2024,Stolpe2024,Abobeih2019}, or spins attached to an external protein/molecule \cite{Schlipf2017}. In such cases, probe-target interactions must be mediated by a transducer, such as the spin network schematically shown in the inset of Fig.~\ref{fig:schematic_illustration}(a) \cite{Stolpe2024,Abobeih2019,Schlipf2017,Fukami2024NVMagnon}. Despite tuning both the network-to-probe and network-to-target interactions to resonance, the significant inhomogeneities typically present within a spin network make it difficult to become polarized, rendering resonant probe-target energy transfer ineffective \cite{Schlipf2017,Stolpe2024,Abobeih2019}.
Here, we instantiate such a scenario by considering the oscillator as the fundamental mode of an NMO, specifically the vibration of a clamped cantilever, as schematically shown in Fig.~\ref{fig:schematic_illustration}(a) \cite{YangNatPhotonics2016,Browaeys2020}.

The paper is organized as follows: In Sec.~\ref{sec:setup}, we outline the Hamiltonian of the considered hybrid system and discuss possible experimental implementations of the proposed protocol. In Sec.~\ref{sec:system_dynamics}, we describe the dynamics of the hybrid system and analyze how periodic inversion pulses on the probe spin can significantly modify it. In Sec.~\ref{sec:cooling}, we detail how repeated probe projections can bring the oscillator and the ensemble into their ground states. In particular, we demonstrate the effectiveness of our protocol by investigating the parameters that allow cooling of the spin ensemble, the oscillator, or both. Additionally, we explore how the same probe control sequence used for cooling can be adapted to prepare the oscillator in complex quantum states.
Finally, Sec.~\ref{sec:dicussion} summarizes our findings and outlines potential avenues for future research.

\section{System Hamiltonian and experimental realization}
\label{sec:setup}

\noindent For simplicity, we assume (though not necessary) that the ensemble spins interact with the common bosonic mode (oscillator) uniformly, with a coupling strength
denoted as \( g \). In addition, the probe spin is assumed to have a distinct
coupling strength with the oscillator, denoted as \( g_0 \). 
The hybrid system's Hamiltonian is then formulated as follows:
\begin{equation}
H = \omega a^{\dagger}a + \sum_{k=0}^N \omega_k S_{k, z} + \left(g_0 S_{0, z} + g\sum_{k=1}^N S_{k, z}\right) \left(a + a^\dagger\right),
\label{eq:initial_system_hamiltonian}
\end{equation}
where $N$ is the total number of spins in the ensemble; \( S_{i,z} \) signifies the z-component of the spin-1/2 operator for the ith spin, oscillating at its Larmor frequency $\omega_i$. The oscillator's annihilation and creation operators are denoted by \( a \) and \( a^\dagger \), with a frequency \( \omega \). 

As we will demonstrate later, the assumption of only a single bosonic mode is valid in this work. This validity stems from our manipulation of the probe spin through periodic inversion pulses, which effectively couples the probe spin exclusively to the oscillator's ground state mode while decoupling it from all others. 
Moreover, our focus on a small ensemble of only a few spins allows for exact diagonalization of the system Hamiltonian. This approach stands in contrast to typical acousto-magnonic studies, where a bosonic description of much larger spin ensembles is necessary \cite{prbrefisart}. Such bosonic approximations are, however, inapplicable in our few-spin limit.

\begin{figure}
    \includegraphics[width=0.48\textwidth]{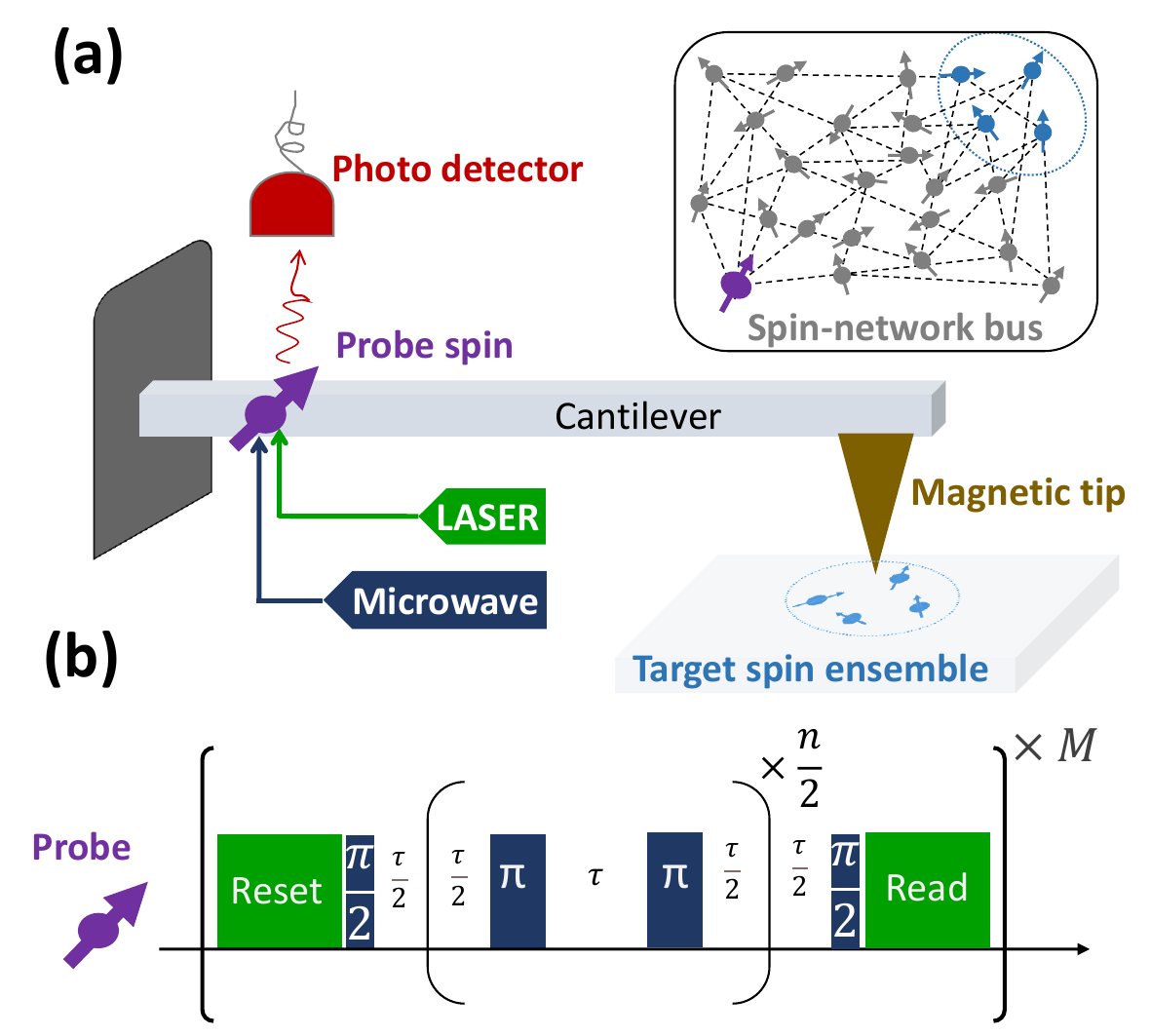}
    \caption{Schematic representation of the experimental setup and control
    strategy. (a) Shown is a clamped nano/micromechanical cantilever carrying a
    single defect spin, which in this study is exemplified by an NV electronic
    spin. The interaction between the NV center spin and the cantilever motion
    is mediated by strain, while single qubit gates are achieved by microwave
    (MW) pulses. Initialization and readout processes are facilitated by optical
    pulses, with a photon detector used for readout. A magnetic tip at the edge
    of the cantilever creates a gradient field that allows interaction with a
    nearby spin ensemble, coupling it to the mechanical motion of the
    cantilever. In the inset, we schematically represent a network of
    interacting spins as a quantum bus. When the number of spins in the network
    is large, it can be modeled as a bosonic system (e.g., magnons) by the
    Holstein-Primakoff transformation. Note that it is simplified here as the
    fundamental mode of the cantilever's vibration. (b) The control sequence of
    the probe spin to perform the cooling protocol. First, the probe spin is
    reset to the $\ket{0}$ state by optical pulses, then a  $\pi/2$ MW pulse
    brings it to a state of equal superposition, which is followed by $n$ inversion MW
    pulses applied periodically with a carefully chosen time interval $\tau$. To ensure that the effective probe-oscillator coupling strength under pulses is comparable to the oscillator frequency \(\omega\), enabling the oscillator to be driven, \(n\) should be on the order of \(\omega / g_0\).
    Another $\pi/2$ MW pulse is applied before the probe is optically read into
    the computational base.  The above process is repeated $M$ times, bringing the
    oscillator and the spin ensemble progressively closer to their thermal
    ground states. Note that the probe to be read must be post-selected in
    $\ket{0}$ each time, otherwise the whole process starts over.}
    \label{fig:schematic_illustration}
\end{figure}

This hybrid system can be realized as shown in Fig. \ref{fig:schematic_illustration}(a) by a unilaterally clamped cantilever, embedding a single NV center positioned at the clamping point such that it experiences maximum strain and thus higher coupling strength. On the free side of the cantilever, we attach a magnetic tip to generate a strong field gradient that couples to a nearby spin ensemble on an external substrate \cite{Rosenfeld2021-ef,Kolkowitz2012Sensing}.  In this way, we exploit both possibilities to achieve spin-mechanical coupling \cite{Rabl2010NPhysics,Bennett2013PRL,Barfuss2015NatPhys} by local (strain) and non-local (magnetic-tip) coupling of spins to mechanical modes within a single setup. This allows the distinct manipulation of quantum systems that are otherwise not controllable. More details about this possible experimental setup can be found in the Appendix.~\ref{sec:setup_details}.

The NV-cantilever setup we've outlined serves as a concrete example of the broader, abstract challenge of harnessing noisy environments. We selected this particular system for two primary reasons. First, achieving polarization is a crucial objective in quantum technology applications using solid-state spins such as NV centers, especially in scenarios where resonant energy transfer is improbable \cite{Sasaki2024PRL,Schwartz2018,Bartling2023a}. Second, and perhaps more significantly, this setup has already been successfully implemented experimentally \cite{loncar,Barfuss2015NatPhys}, providing a solid foundation for further investigation of using noisy environment as resources.

\section{System dynamics and its tuning}
\label{sec:system_dynamics}

\noindent In this section, we provide analytical derivations of the system dynamics and illustrate how periodic inversion pulses on the probe spin can significantly affect the behavior of the whole system when certain resonance conditions are roughly met. These derivations are intended to provide insight into the mechanisms of our cooling protocol shown schematically in Fig.~\ref{fig:schematic_illustration}(b).

\subsection{Original system dynamics}

\noindent In the following, we will consider the system in the rotating frame with respect to the precession frequencies of both the oscillator and the spins. This transformation is applied to an arbitrary quantum state $\ket{\psi}$ as follows: $\exp{(-i\omega a^{\dagger}a)}\exp{(-i\sum_k \omega_k S_{k,z})}\ket{\psi}$ \cite{Wang2023PhdThesis}. Then the rotating-frame Hamiltonian reads
\begin{equation}
    \tilde{H} = \left(g_0 S_{0, z}+ g\sum_{k=1}^N S_{k, z}\right) \left(\hat{a}e^{-i\omega t} + \hat{a}^{\dagger}e^{i\omega t}\right).
\end{equation}
This Hamiltonian allows us to study the effect of our cooling protocol on each spin configuration with a given eigenvalue of the collective spin operator $S^z =\sum_{k=1}^N S_{k, z}$, independently. The system evolution operators are then conveniently decomposed into the following form \cite{Agarwal2010-rf,Rao2016-uw}:
\begin{equation}
    \begin{split}
        U(t) &= \ket{0}\bra{0}\otimes \sum_{\scriptscriptstyle k = 0}^{\scriptscriptstyle N} \mathcal{D}_{k, +}(t)\otimes\mathrm{I}_k \\
        &\qquad \qquad  + \ket{1}\bra{1}\otimes\sum_{\scriptscriptstyle n = 0}^{\scriptscriptstyle N}\mathcal{D}_{n, -}(t)\otimes\mathrm{I}_k,\\
        &\mathcal{D}_{k, \pm}(t) = \mathcal{T} \exp\left(\pm i\int_0^t \mathrm{d}t'\ h_{\scriptscriptstyle k, \pm}(t')  \right),
    \end{split}
    \label{eq:DD_modulated_evolution}
\end{equation}
where $\mathcal{T}$ the time ordering operator; the terms $h_{\scriptscriptstyle k, \pm}(t)$ are the oscillator Hamiltonians conditioning on the spin states, which are expressed as:
\begin{equation}
    h_{\scriptscriptstyle k, \pm}(t) = \frac{g(2k-N) \pm g_0}{2} \left(\hat{a}e^{-i\omega t} + \hat{a}^{\dagger}e^{i\omega t}\right);
\label{eq:hk_pm}
\end{equation}
and the operator $\mathrm{I}_{\scriptscriptstyle k}$ projects the spin ensemble into the subspace where $k$ spins are pointing up. 
For example, when $N=3$ and $k = 2$, the projector is written as
\begin{equation}
    \mathrm{I}_{\scriptscriptstyle k=2} = \ket{\uparrow \uparrow \downarrow}\bra{\uparrow \uparrow \downarrow} +  \ket{\uparrow  \downarrow \uparrow}\bra{\uparrow \downarrow \uparrow} + \ket{\downarrow \uparrow \uparrow }\bra{\downarrow \uparrow \uparrow}.
    \label{eq:ensemble_projector}
\end{equation}
It has been shown that the Magnus expansion of the evolution operators can be simplified as \cite{Agarwal2010-rf,Rao2016-uw}
\begin{equation}
    \begin{split}
    \mathcal{D}_{\scriptscriptstyle k, \pm}(t&)  = \exp\left(\pm i\int_0^t \mathrm{d}t' h_{\scriptscriptstyle k, \pm}(t')\right)\\
    & \qquad \times \exp\left(\frac{1}{2}\int_0^t \mathrm{d}t'\int_0^{t'} \mathrm{d}t'' \left[ h_{\scriptscriptstyle k, \pm}(t'), h_{\scriptscriptstyle k, \pm}(t'')\right]\right)\\
    & = \exp\left[\frac{g(2k-N)\pm g_0}{2\omega}\left(\alpha(t) a^\dagger-\alpha^*(t) a\right)\right] \\
    &\qquad \quad \times \exp\big(it\theta_{\scriptscriptstyle k}\big).
    \end{split}
    \label{eq:conditional_evol_op_noDD}
\end{equation}
where  $\alpha(t) = 1-e^{i\omega t}$ and $\theta_{\scriptscriptstyle k}$ is given by
\begin{equation}
\begin{split}
    \theta_{\scriptscriptstyle k} &= -\frac{\omega t - sin(\omega t)}{\omega t}\frac{\left[g(k-N/2)\pm g_0\right]^2}{4\omega}.
\end{split}
\end{equation}
The analysis can be significantly simplified by considering each spin component with a fixed value of $k$ separately. The initial ensemble state is diagonal in the basis of the collective spin operator $S^z$, and this diagonal nature is preserved throughout the complete system evolution. Consequently, each $k$-dependent angle $\theta_k$ becomes a global phase for the corresponding spin component with a particular eigenvalue of $S^z$, allowing us to disregard it in our analysis. 
It is also important to note that the oscillator-induced spin-spin interaction,  plays no role in the cooling process, as it also takes the form $\propto \sum_{i,j}{S_i^zS_j^z}$ \cite{RaoPRL2011}.

\subsection{Dynamical Decoupling pulses on probe}
\noindent The original system dynamics are unlikely to exhibit energy transfer-like effects due to the weak and off-resonant spin-oscillator couplings. To introduce some tunability into these dynamics, we consider applying a periodic sequence of \(\pi\) pulses to the probe spin. Our cooling method has some similarity to how high-fidelity entangling gates can be realized between an NV electronic spin and a nuclear spin, which is also achieved by periodic inversion pulses on the electronic spin \cite{Taminiau2014}.

Similar to the nulear-spin polarization schemes \cite{Wang2023PhdThesis,Bradley2019}), our cooling protocol relies on significantly altering the system dynamics by fine-tuning the pulses on the probe spin to match the oscillator freqeuncy. This allows the effects of weak spin-oscillator couplings to accumulate over time, thereby enabling effective cooling of the spin ensemble and the oscillator.
These pulses alter the dynamics of the system by replacing $g_0$ with $f(t)g_0$ in Eq.~(\ref{eq:hk_pm}). The function $f(t)$ is time-periodic and expressed as
\begin{equation}
    f(t)=
        \begin{cases}
            +1 & \text{if } 2j\tau< t < (2j+1)\tau\\
            -1 & \text{if } (2j+1)\tau< t < (2j+2)\tau
        \end{cases},
\end{equation}
where \( j = 0, 1, 2, \cdots, n\) and $\tau$ denotes the interval between successive pulses. The conditional evolution operators in Eq.~(\ref{eq:conditional_evol_op_noDD}) are now written as
\begin{equation}
    \begin{split}
        \mathcal{D}^f_{\scriptscriptstyle k, \pm}(t) = &\exp\Big[\frac{g(k-N/2)}{2\omega}\left(\alpha(t)a^{\dagger} - \alpha^*(t)a\right) \\
        & \; \pm \frac{g_0}{2\omega}\left(F(\epsilon, t)a^{\dagger} - F^*(\epsilon, t)a\right)\Big]\times \exp\left[i\theta_{\scriptscriptstyle k}(t)\right],
    \end{split}
    \label{eq:equation9}
\end{equation}
where the filter function $F(t)$ is given by \cite{Agarwal2010-rf}
\begin{equation}
    \begin{split}
        F(t) &= i \omega\ \sum_{k=0}^{n}\int\limits_{k\tau}^{(k+1)\tau}\mathrm{d}t\;(-1)^k e^{-i\omega t}\\
        &=\left[1 - (-1)^{n+1}e^{-i(n+1)\omega\tau} + 2\sum_{k=1}^{n}(-1)^k e^{-ik\omega\tau}\right].
    \end{split}
\end{equation}
This filter function is centered at $\tau = \pi/\omega$ with its width decreasing approximately as $1/n^2$ \cite{Rao2016-uw, Agarwal2010-rf, Uhrig2007-nu} and asymptotically converges to a delta function:
\begin{equation}
    \lim\limits_{n\rightarrow \infty} F(t) = 2(n+1)\delta(\omega - \frac{\pi}{\tau}),
    \label{eq:filter_in_limit}
\end{equation}
which supports our initial assumption that all other phonon modes are effectively decoupled in Eq.~(\ref{eq:initial_system_hamiltonian}) with large number of pulses $n$.
In addition, we also want the maximum possible coupling strength between them. One can see from Eq.~(\ref{eq:equation9}) that $|g_0 F(t)|$ should be comparable to the oscillator frequency $\omega$, which requires at least $n \approx \mathcal{O}(\omega/g_0$) \cite{Rao2016-uw}. 

The average displacement magnitude caused by the ensemble spins can be approximated as $\alpha_E(k) \approx g(k-N/2)/2\omega$, which is notably small compared to the probe-induced displacement. The latter, given by $\alpha_P \approx ng_0/\omega$, scales with the number of pulses $n$ when the resonance condition is met, as shown in Eq.~(\ref{eq:filter_in_limit}). Consequently, in a free induction decay experiment of the probe spin, where no inversion pulses are applied, the ensemble-induced effects on the oscillator remain undetectable. However, in our protocol, each probe projection modifies the oscillator state, conditioning it on these minute ensemble-induced displacements. Although initially negligible, these effects accumulate over the repeated probe projections, ultimately exerting a significant influence on the system dynamics.

\begin{figure}

\includegraphics[width=0.48\textwidth]{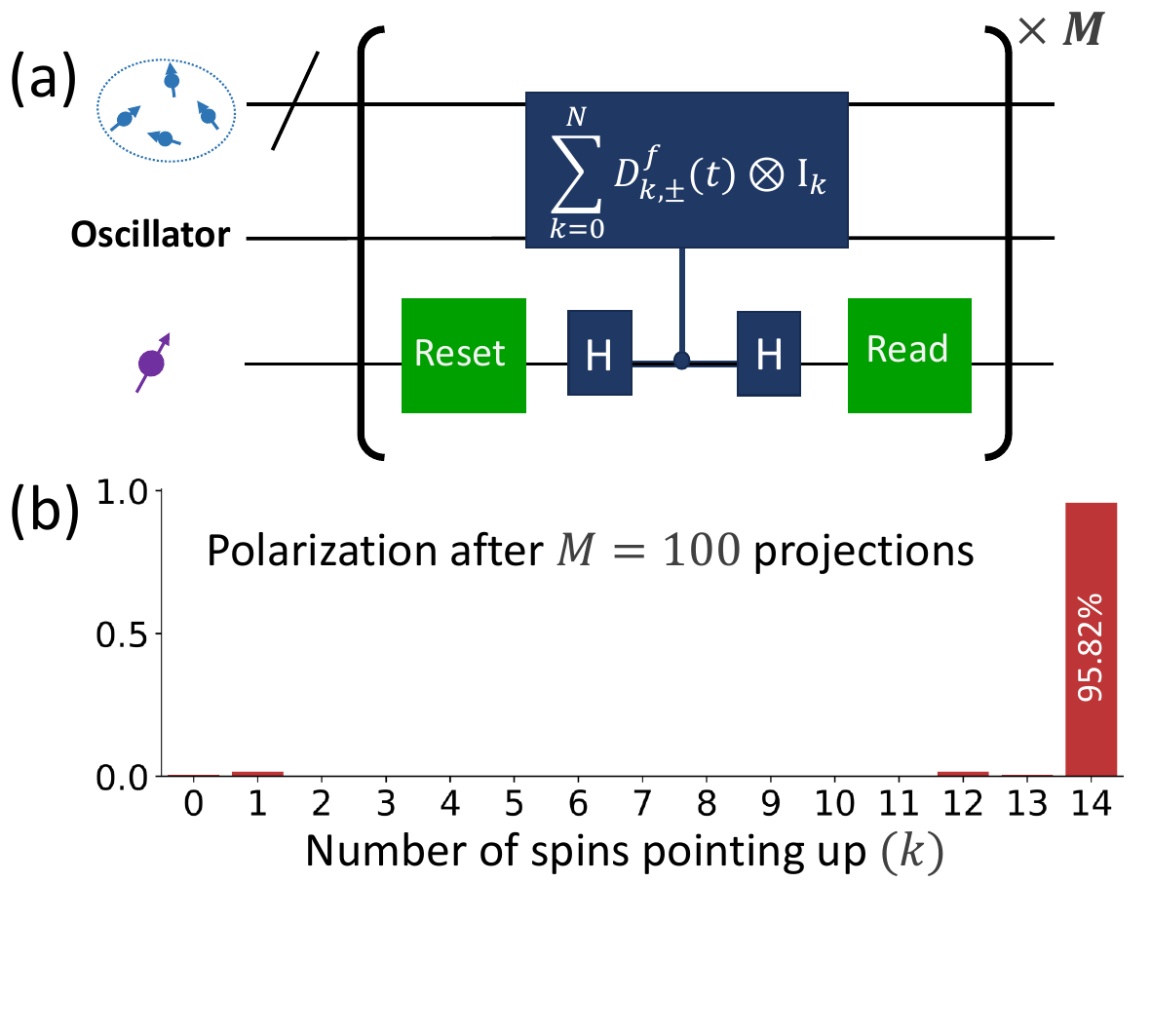}
\vspace{-15mm}
    \caption{(a) The quantum circuit progressively projects the oscillator ensemble system by projection of the probe spin. The pulse sequence on the physical level is shown in fig.~\ref{fig:schematic_illustration}(b). In each repetition, the probe spin is first prepared in the equal superposition state \(\ket{+}\) by the optical reset and a microwave Hadamard gate.  Then repeated inversion pulses are applied, resulting in a controlled oscillator displacement \(\sum_k \mathcal{D}^f_{\scriptscriptstyle k, \pm}(t)\otimes \mathrm{I}_k\).  Here the subscript $\pm$ depends on the spin state of the probe and $k$ on the number of upward-pointing ensemble spins. The procedure ends with a Hadamard gate and a measurement of the probe spin, with the result set to 0. Such a repetition effectively implements the projector given in Eq.~(\ref{eq:oscillator_ensemble_projector}). (b) An example of how an ensemble of $N=14$ spins is polarized after $M=100$ probe projections. The spins are initially in a completely mixed state, which is projected to the state with all spins pointing up with a probability of 95.82\%. Note that achieving such a high degree of polarization requires a carefully chosen pulse detuning, which in this case is $\epsilon=1.16$ kHz. Further simulation details are summarized in Appendix~\ref{sec:simul_detail}. }
\label{fig:ancilla_ensemble_dependent_oscillator_displacement}
\end{figure}

\subsection{Tuning dynamics through probe projections}
\noindent 
In this section, we show how the system dynamics can be modulated by applying inversion pulses to the probe spin. 
The dynamics modification is facilitated by a cyclic process involving projective measurement and reinitialization of the probe spin, as schematically shown in Fig.~\ref{fig:schematic_illustration}(a). 

First, the probe is prepared in the state \(\ket{+}\). Then the system undergoes an evolution for a time \(t = n \tau\) under repeated inversion pulses as described in Eq~(\ref{eq:DD_modulated_evolution}). Finally, the probe spin is measured in the computational basis after an additional \(\pi/2\) pulse is applied to it. 
This process is represented by the circuit in Fig.~\ref{fig:ancilla_ensemble_dependent_oscillator_displacement}(a), which effectively implements oscillator displacement conditioning on the probe spin state. 

The whole circuit is equivalent to the implementation of projection operators on the oscillator that depend on the probe spin as well as the ensemble spin states. These conditional oscillator projection operators \(V_{k, \pm}\) are mathematically formulated as:
\begin{equation}
    \mathcal{P}_{\scriptscriptstyle N, \pm} = \sum_{\scriptscriptstyle k = 0}^{\scriptscriptstyle N}V_{k, \pm} \otimes \mathrm{I}_k, \quad V_{k, \pm} = \frac{1}{2} \left(\mathcal{D}^{\scriptscriptstyle f}_{\scriptscriptstyle k, +}(t) \pm \mathcal{D}^{\scriptscriptstyle f}_{\scriptscriptstyle k, -}(t)\right),
    \label{eq:oscillator_ensemble_projector}
\end{equation}
where \(k\) is the number of ensemble spins pointing up, and the subscript \(\pm\) depends on the measurement result of the probe spin. Such repeated projections of the probe are central to the oscillator cooling protocol in our previous work \cite{Rao2016-uw}, which is now extended to cooling/polarizing a spin ensemble. 

We start with an ensemble of \( N \) spins initially in a fully mixed state and the oscillator initially in its thermal state. These initial states are written as
\begin{equation}
    \rho_{\text{ens}} = \sum_{k=0}^{N} \frac{\mathrm{I}_k}{2^N}, \quad \rho_{\text{osc}} = \frac{1}{\sum_n e^{-\frac{n\omega}{k_{\scalebox{0.5}{$B$}} T}}}\sum_n e^{-\frac{n\omega}{k_{\scalebox{0.5}{$B$}}T}}\ket{n}\bra{n},
\end{equation}
where projector \(\mathrm{I}_k\) corresponds to the states with 
\(k\) spins pointing up, as given in Eq.~(\ref{eq:ensemble_projector}); \( k_{\scalebox{0.5}{$B$}} \) represents the Boltzmann constant; and \( T \) is the temperature. It is easy to see that \( \langle \sum_{j=1}^N S_{z, j} \rangle = 0 \); and the initial thermal occupancy of the oscillator is calculated as $n_0 =\Tr(\rho_{osc}\, a^\dagger a) = 1/\left[\exp{(\omega/k_{\scalebox{0.5}{$B$}}T)}-1\right]$.

To cool the ensemble-oscillator subsystem, we repeatedly implement the circuit shown in Fig.~\ref{fig:ancilla_ensemble_dependent_oscillator_displacement}(a) by post-selecting the probe measurement outcome to be 0 each time. This process effectively implements the projector $\mathcal{P}_{\scriptscriptstyle N, +}$ given in Eq.~(\ref{eq:oscillator_ensemble_projector}) many times. The probability that the ensemble is projected into the $\mathrm{I}_k$ sector after $M$ repetitions is calculated as
\begin{equation}
    \mathrm{P}_m(M) =\frac{1}{\sum_m \mathrm{P}_m } \mathrm{Tr}[\mathcal{P}_{\scriptscriptstyle N, +}^M\,\rho_B(0)\,\mathcal{P}_{\scriptscriptstyle N, +}^M],
    \label{eq:final_projection_portion}
\end{equation}
where $m=(2k-N)/2$ is the total magnetization of the ensemble. This dependence on $m$ shows that, with appropriate parameters, it is possible to polarize the ensemble, i.e., to obtain $\mathrm{P}{\scriptscriptstyle N/2}(M) \approx 1$ when $M$ is large enough.

\subsection{Pulse detuning}
\noindent To achieve tunability of the system dynamics described in Eq.~(\ref{eq:final_projection_portion}) and facilitate our proposed cooling protocol, we introduce the pulse detuning $\epsilon$ as a control parameter:
\begin{equation}
    \tau = \frac{\pi}{\omega - \epsilon}, \text{ with } \frac{\epsilon}{\omega} \ll 1.
    \label{eq:resonance_condition}
\end{equation}
Note that the detuning is intentionally kept significantly smaller than the oscillator frequency, so that the inversion pulses are in the near-resonance regime.
Accordingly, the filter function can now be simplified as
\begin{equation}
        F(\epsilon, t = (n+1)\tau) = 1 - e^{i\epsilon t} + \frac{2(1-e^{i\epsilon t})}{e^{-i\epsilon \tau}-1}.
        \label{eq:filter_function}
\end{equation}
For higher values of pulse detuning $\epsilon$, making the above equation closer to the delta function given in Eq.~(\ref{eq:filter_in_limit}) would require an even larger pulse number $n$.

\section{System cooling}
\label{sec:cooling}
\noindent 
In this section, we demonstrate the ability of repeated probe projections to impose a thermal filter on the hybrid system, allowing controlled cooling of either the oscillator, the ensemble, or both. In the case of simultaneous cooling, the oscillator can only be cooled after the ensemble is fully polarized; otherwise, there is effectively no interaction between the probe and the ensemble. As illustrated in Fig.~\ref{fig:simultaneous_cooling}, the cooling process for the entire system is divided into two stages with distinct pulse spacings \(\tau\). In the first stage, the oscillator remains in its thermal state while the ensemble is cooled. In the second stage, another pulse spacing is used to rapidly cool the oscillator, following the oscillator cooling scheme described in our previous work \cite{Rao2016-uw}.

We explore a range of parameters to identify this effect through numerical simulations, focusing primarily on the weak coupling regime where the coupling strength \(g\) is significantly smaller than the mechanical oscillation frequency \(\omega\), denoted as \(g \ll \omega\). Details about these simulations are summarized in Appendix~\ref{sec:simul_detail}.

To explore this weak-coupling regime, we set the spin coupling strengths \(g, ~ g_0\) to \(10\) kHz, while the mechanical oscillation frequency \(\omega\) is chosen to be \(1.2\) MHz. Although such a frequency for the mechanical oscillator is experimentally achievable \cite{Verbridge2008,Li2007}, the chosen values for the spin-mechanical couplings are at the upper end of what is typically achievable in practice \cite{Rao2016-uw}. 

\begin{figure}[htbp!]
\includegraphics[width=0.48\textwidth]{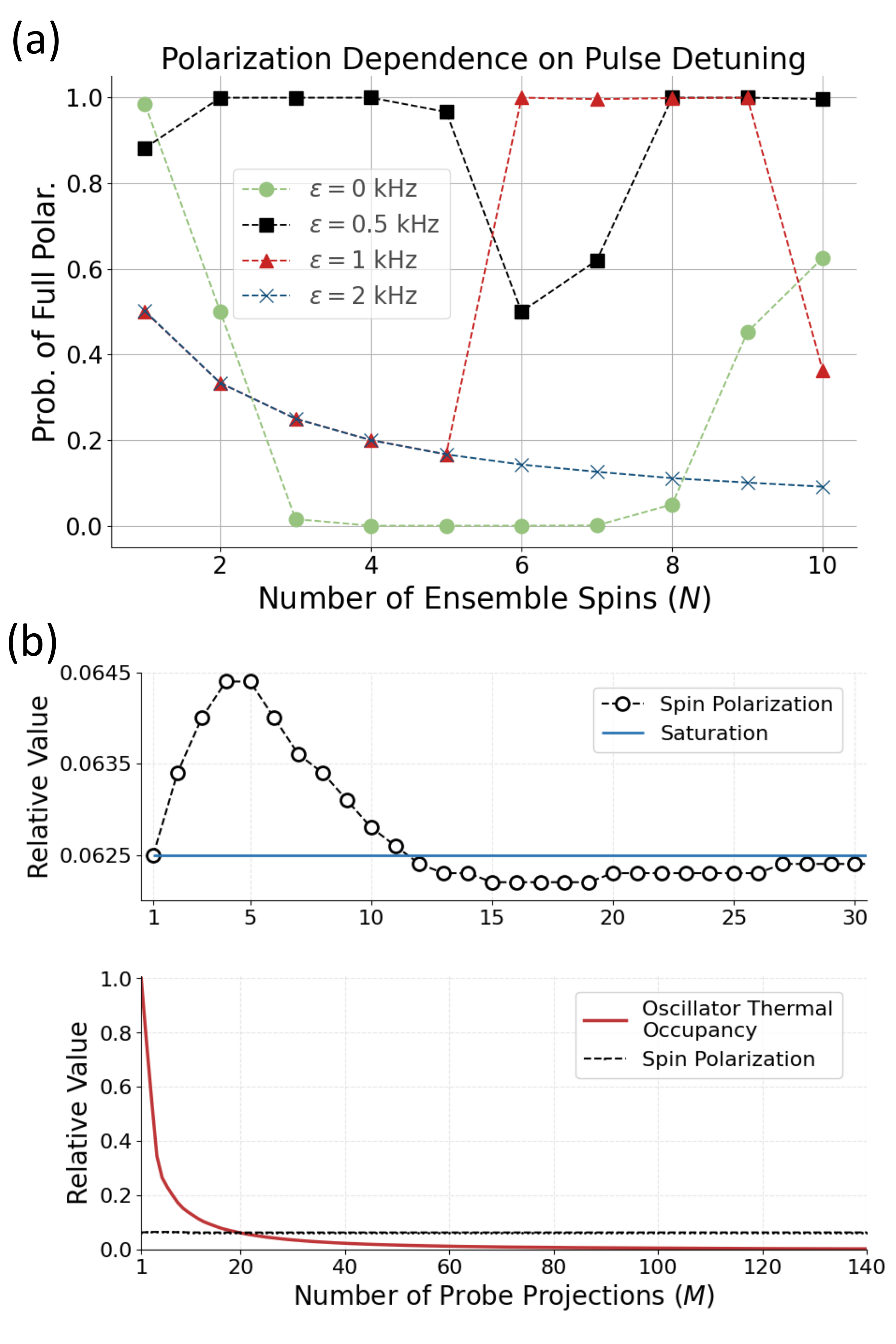}
    \caption{Pulse Detuning: A key parameter in the cooling scheme. (a) This panel shows the intricate dependence of achieving complete spin polarization on pulse detuning (\(\epsilon\)). The probability that all ensemble spins are aligned upward after \(M=100\) probe projections, denoted by \(\mathrm{P}_{\scriptscriptstyle m=N/2}(M)\) and derived in Eq.~(\ref{eq:final_projection_portion}), is plotted on the vertical axis. This probability is affected by both the number of spins in the ensemble and the size of \(\epsilon\), with \(\epsilon = 2\) kHz already appearing too large for effective polarization of the ensemble. (b) If we choose an even larger detuning (\(\epsilon=2.5\) kHz) and set the number of spins to \(N=4\), we observe a rapid decrease of the thermal occupancy of the oscillator to zero, indicating that the oscillator is cooling. However, the spin polarization remains almost unchanged and reaches a steady state of 0.0625 after about 40 probe projections. Note that we have considered an ensemble of $N=4$ spins, so the y-axis represents the relative value of $\mathrm{P}_z(M)/2$. Correspondingly, the y-axis also represents the relative value of the thermal occupancy of the oscillator with respect to its initial value, which is set to about 45. Further simulation details are summarized in Appendix~\ref{sec:simul_detail}. }
    \label{fig:probabilities_of_projections}
\end{figure}

These choices represent a compromise between stronger coupling for faster cooling and additional noise due to imperfect pulses on the probe.  As we discussed in Eq.~(\ref{eq:filter_in_limit}), at least \(\mathcal{O}(\omega/g_0)\) pulses are required for each probe projection to ensure that the effective spin-oscillator coupling is comparable to the oscillator frequency, which becomes clear in Eq.~(\ref{eq:equation9}). However, a higher number of pulses would result in a longer total evolution time to implement the cooling protocol, which is undesirable in the presence of decoherence. Throughout this work, we set the number of pulses in each round of probe projection to \(n=\omega/g_0 = 120\),  ensuring that the effective probe-oscillator coupling strength is sufficiently strong to drive the oscillator, as demonstrated in Eq.~(\ref{eq:equation9}).

In this work, we refrain from exhaustively exploring all possible parameter regimes. Instead, we focus on fixed coupling strengths and oscillator frequencies that are experimentally feasible, and consider only a few spins in the ensemble. Our primary goal is to demonstrate that manipulating the probe spin, specifically by tuning the pulse detuning amplitude \(\epsilon\), can significantly alter the dynamics of a hybrid oscillator-ensemble system with only weak dispersive couplings. The parameter regimes we explore are sufficient to confirm this capability, as we will discuss below.

\subsection{Pulse detuning as a control parameter}

\noindent 
The probability $\mathrm{P}_m(M)$ in Eq.~(\ref{eq:final_projection_portion}) depends non-trivially on the filter function in Eq.~(\ref{eq:filter_function}) that is imposed by pulses on the probe.
As discussed in our previous work \cite{Rao2016-uw}, varying the value of $\epsilon$ can lead to cooling, heating, or squeezing of the oscillator. 
In particular, Ref.~\cite{Rao2016-uw} has found that a ratio of $\epsilon/\omega \approx 5 \times 10^{-3}$ leads to rapid cooling of the oscillator, which corresponds to $\epsilon \approx 6$kHz for the oscillator frequency of $1.2$ MHz.

However, to achieve complete polarization of the ensemble, we must avoid premature cooling of the oscillator before the ensemble is polarized. This is due to the significant effect of the thermal occupancy of the oscillator on the indirect interaction strength between the probe and the ensemble, as elaborated in Eq.~(\ref{eq:initial_system_hamiltonian}). In particular, at zero thermal occupancy, the oscillator-mediated effective interaction between the probe and the ensemble spins goes correspondingly to zero.

First, we run simulations with the pulse detuning \(\epsilon\) close to zero to explore the near-resonance regime. By adjusting its magnitude from 0 to 2 kHz and varying the number of spins in the ensemble, our simulations reveal a nuanced relationship between the probability of achieving complete ensemble polarization and the pulse detuning \(\epsilon\), as shown in Fig.~\ref{fig:probabilities_of_projections}(a). 
To cool the spin ensemble, the choice of $\epsilon$ should satisfy the condition that the probe-oscillator coupling is enhanced without prematurely cooling the oscillator. Consequently, satisfying this condition requires fine-tuning of all parameters of the Hamiltonian. 

This makes it difficult to determine the exact dependence of the optimal pulse detuning $\epsilon$ on the number of ensemble spins $N$ and the number of probe projections $M$, as shown in Fig.~\ref{fig:probabilities_of_projections}(a). 
For example, it can be seen that within an ensemble of constant size, varying the detuning can lead to very different results when implementing the proposed cooling protocol.
Another important observation from Fig.~\ref{fig:probabilities_of_projections}(a) suggests that a detuning of \(\epsilon=2\) kHz may be large enough to facilitate rapid oscillator cooling and thereby prevent ensemble polarization.

\subsection{Premature oscillator cooling}

\noindent To investigate the premature cooling of the oscillator further, we introduce an even larger detuning of \(\epsilon=2.5\) kHz for an ensemble of $N=4$ spins. The simulation results shown in Fig.~\ref{fig:probabilities_of_projections}(b) indicate that at this substantial pulse detuning, the oscillator quickly reaches its ground state after about 80 probe projections. This premature cooling of the oscillator reduces the interaction between the probe and the ensemble, so that the ensemble remains largely unpolarized. The ensemble polarization fluctuates only slightly around the final saturation value for the first few tens of probe projections, as shown in the inset of Fig.~\ref{fig:probabilities_of_projections}(b).

Furthermore, as an example, we search for the optimal pulse detuning for cooling a larger ensemble of $N=14$ spins. An exhaustive parameter search leads us to set the detuning to $\epsilon = 1.16$ kHz. This setting shows that the sector with all spins pointing up (i.e. \(m=7\)) is predominantly left after $M=100$ probe projections. The probability of full polarization is as high as 95.82\%, as shown in Fig.~\ref{fig:ancilla_ensemble_dependent_oscillator_displacement}(b).

The above results underscore the critical role of \(\epsilon\) as a controlling parameter in the cooling protocol. In addition, a subtle and interesting aspect of cooling the oscillator is to choose the detuning $\epsilon$ precisely so that the conditional oscillator displacement operator is not perfectly aligned with the position or momentum quadrature. Otherwise, the oscillator will be squeezed instead of cooled \cite{Rao2016-uw}. Note that, in the resonant case that $\epsilon=0$, the displacement operator of the oscillator is along the position quadrature.

This observation is also interestingly related to the encoding of Gottesman-Kitaev-Preskill (GKP) states, which have applications in quantum sensing and quantum error correction \cite{Gottesman2001-tl,Terhal2016-sy,Vuillot2019_qec}. Its encoding can be realized by the same control of the probe spin as our cooling protocol. Starting with an oscillator in its zero photon ground state, repeated probe projections can actually increase the number of photons and leave the oscillator in GKP states \cite{Gottesman2001-tl,Terhal2016-sy}.
The generation of these states in macroscopic objects could enrich our understanding of the quantum-classical interface, notwithstanding their important roles in quantum error correction \cite{Gottesman2001-tl,Terhal2016-sy} and quantum sensing \cite{Kasper2017_sensor}. More details of GKP encoding can be found in  Appendix~\ref{sec:GKPEncoding}.

\begin{figure}
\vspace{-5mm}
\includegraphics[width=0.48\textwidth]{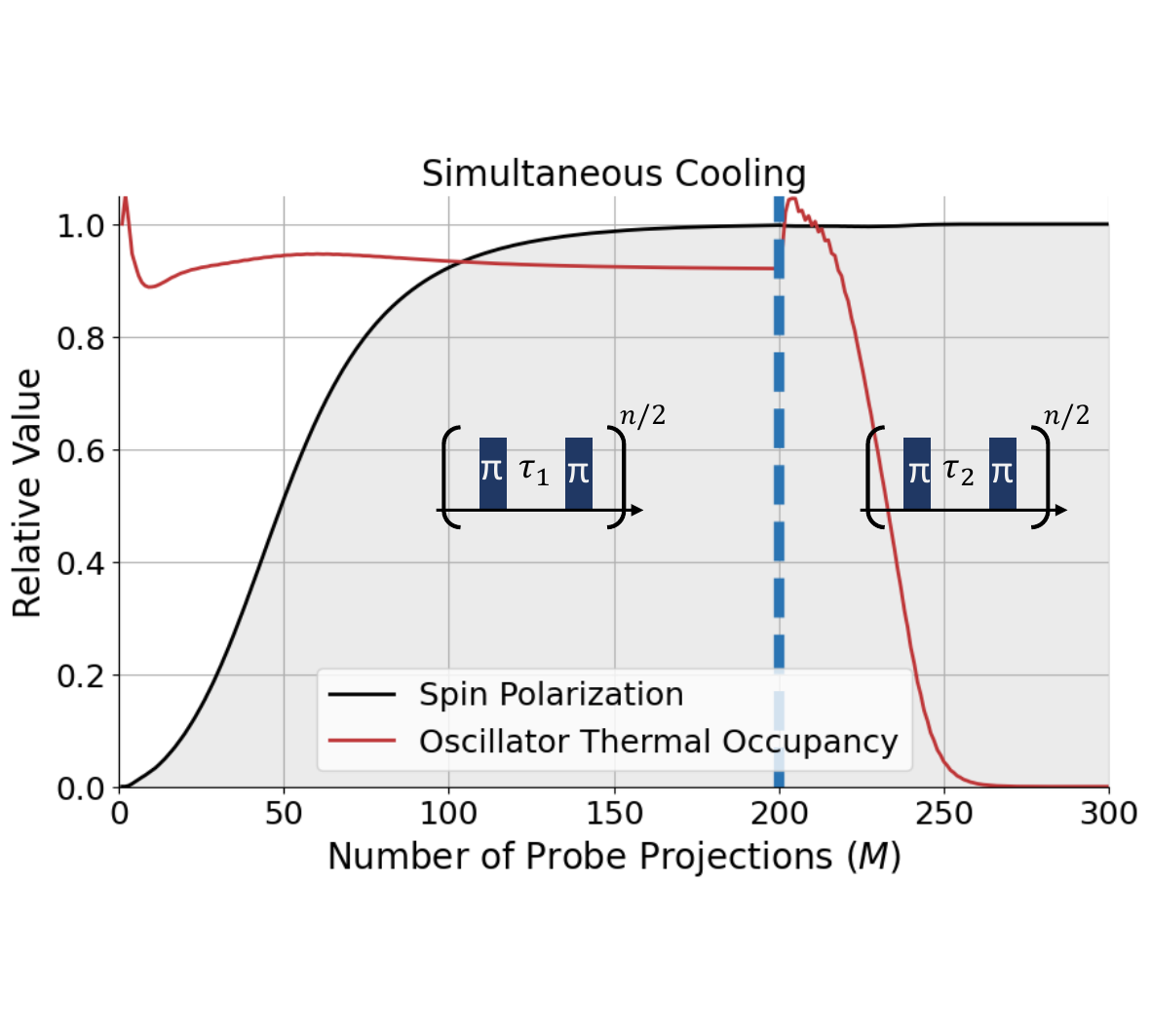}
\vspace{-10mm}
    \caption{Cooling of both the oscillator and the ensemble. This figure shows the polarization $\mathrm{P}_z(M)$ and the thermal occupancy $n(\omega)$ for both the spin ensemble and the oscillator, illustrating how the cooling/polarization proceeds with the number of probe projections $M$. The effective cooling of these quantum systems occurs in different parameter regimes. Initially, the spin ensemble is cooled by adjusting the pulse intervals to near-resonance conditions, specifically $\tau_1$ such that $\epsilon_1 = 1.1$ kHz. In this case, the thermal occupancy of the oscillator remains largely unaffected. After polarizing the spin ensemble, we change the pulse interval to $\tau_2$ with a pulse detuning of $\epsilon_2 = 5$ kHz. This more significant detuning leads to a rapid cooling of the oscillator to its ground state. Note that the number of ensemble spins is $N=4$ so that the y-axis represents the relative value of spin polarization $\mathrm{P}_z(M)/2$. The oscillator thermal occupancy is also plotted as its relative to the initial occupancy $n_0\approx 45$. Further simulation details are summarized in Appendix~\ref{sec:simul_detail}.}
    \label{fig:simultaneous_cooling}
\end{figure}

\subsection{Simultaneous cooling of ensemble and oscillator}

\noindent 
In this section, we discuss the cooling of both the spin ensemble and the oscillator simultaneously. The key to achieving this goal is to avoid premature cooling of the oscillator, which should instead occur after the ensemble is fully polarized. This requires tuning the pulse detuning  \(\epsilon\) separately for the cooling phases of the oscillator and the ensemble.
Fig.~\ref{fig:simultaneous_cooling} illustrates a pulse sequence that effectively polarizes an ensemble of \(N=4\) spins while cooling an oscillator from an initial thermal occupancy of \(n_0 \approx 45\).

During this simulation, we find that the optimal detuning that achieves complete polarization of the ensemble is identified as \(\epsilon_1 = 1.1\)kHz, while the oscillator occupancy is not much reduced. In this case, performing 200 probe projections can achieve complete polarization of the ensemble. Then, to cool the oscillator, we adjust the detuning to a higher value, \(\epsilon_2 = 5\)kHz, and execute an additional series of \(50\) probe projections. This transition in the magnitude of the pulse detuning results in a rapid cooling of the oscillator, as shown in the right part of Fig.~\ref{fig:simultaneous_cooling}.

It is important to note that increasing \(\epsilon_2\) requires a higher number of inversion pulses on the probe to achieve a comparable cooling effect, as can be seen from Eq.~(\ref{eq:filter_function}). However, implementing a higher number of inversion pulses presents experimental challenges, primarily due to the potential for pulse imperfections. This consideration emphasizes the need for precise control and optimization in the experimental setup.

\subsection{Effects of decoherence}

\label{sec:decoherence}
\noindent 
Our numerical simulations have not considered the direct effect of spin dephasing noise. For the hybrid system considered in this work, such noise is induced by the spin bath surrounding the probe and the ensemble and is typically the dominant noise source \cite{Wang2023PhdThesis,Abobeih2018}. Firstly, the repeated inversion pulses effectively protect the probe spin from dephasing noise. Notably, this technique can extend the coherence time of an NV electronic spin to exceed one second under cryogenic temperatures \cite{Abobeih2018}. Consequently, such a long coherence time is not a limiting factor for our protocol. Additionally, after each probe projection, both the spin ensemble and its surrounding spin bath gradually collapse towards their steady states simultaneously, a process that is sensitive only to the probe population in its measurement basis \cite{durgancommn}. One thing worth noting is that previous works, such as Refs.~\cite{Rao2020NJP, Sasaki2024PRL}, have counterintuitively shown that decoherence of the spins in an ensemble effectively enhances its polarization rather than decreasing it.

Similarly, regarding the oscillator cooling, the effects of the oscillator dissipation (\(T_1\)) is also negligible. As shown in Fig.~\ref{fig:simultaneous_cooling}, the initial bosonic occupancy of the oscillator remains almost unaffected during ensemble cooling. Therefore, the effect of the oscillator's \(T_1\) on ensemble cooling can be ignored. Once the ensemble is fully polarized, there is only one spin configuration with all spins in the same state. The spin ensemble now exerts only a static force on the oscillator, displacing the oscillator to a new equilibrium. The right panel of Fig.~\ref{fig:simultaneous_cooling} hence exactly corresponds to the oscillator cooling scheme presented in our previous work \cite{Rao2016-uw}, where it was shown that the $(T_1)$ decay of the oscillator only influences the total number of probe projections required to achieve a similar cooling effect i.e., $T_1$ decreases the cooling rate but not the achievable cooling limit.

We note that other imperfections, such as inhomogeneous ensemble-oscillator coupling and imperfect probe readout, will affect the cooling efficiency but not the final state that can be achieved. A more detailed discussion can be found in the Appendix.

\subsection{Cooling efficiency}

\noindent It is crucial to note that our protocol only accepts a probe measurement result of 0; any other outcome necessitates restarting the entire process. This requirement leads to an exponentially decreasing success rate, which constitutes the primary limitation in achieving full polarization of a spin ensemble in practice. In contrast, the relatively weaker ensemble-oscillator coupling strength ($g < g_0$) does not qualitatively alter our cooling protocol. However, it does affect the cooling rate, necessitating more probe spin projections to achieve the desired level of cooling. While these limitations make achieving full polarization with our protocol challenging, demonstrating partial polarization is expected to be feasible and would already constitute an exciting experiment.

Moreover, traditional methods, such as those based on resonant energy transfer, can initially be used to achieve a substantial degree of polarization in a spin ensemble. Our protocol can then be applied to further enhance the polarization to levels otherwise unattainable with energy transfer alone. This is because our protocol essentially projects out components with lower polarization, thereby increasing the success rate of obtaining desired probe measurements as the spin ensemble becomes more polarized. In the extreme case of an almost fully polarized spin ensemble, the success probability of a probe projection would be close to 1. 

\section{Discussion}
\label{sec:dicussion}

\noindent  In this study, we have developed a novel ground-state cooling protocol for a tripartite hybrid quantum system consisting of a spin ensemble, an oscillator, and a probe spin. Central to this protocol is the exploitation of the feedback provided by frequent measurements of the prob spin. Our analysis has shown that even with weak dispersive coupling between the spins and the oscillator, one can effectively cool both the ensemble and the oscillator to their ground states, either separately or simultaneously. Our analysis has also revealed a more complex relationship between the achievable polarization levels and the ensemble size. This relationship is further complicated by the detuning $\epsilon$, making it challenging to derive a straightforward scaling behavior for the required number of projections $M$ as a function of the ensemble size $N$. Establishing such scaling relations requires a more detailed analysis, which we leave it for future study.

Our cooling protocol offers several desirable features. First, it requires quantum coherence only in the probe, distinguishing it from resonant transfer schemes that demand coherence throughout the hybrid system. In addition, the periodic inversion pulses significantly enhance the probe spin's protection, extending its coherence time well beyond current technological requirements. Furthermore, when applied to a macroscopic oscillator, this polarization technique can be viewed as a remote sensing scheme. This approach allows the sensing volume boundaries to extend beyond the limits imposed by phase accumulation schemes \cite{Staudacher2013}, offering potential advantages in various sensing applications.

Numerical simulations are performed over a range of parameters, underscoring the feasibility of the cooling protocol and laying the groundwork for its experimental realization. For example, our results indicate that a spin ensemble consisting of \( N = 14 \) spins could be cooled with a spin-mechanical coupling strength of only a few kHz at a mechanical oscillation frequency close to 1 MHz. Although such coupling strengths are within experimental reach, the cooling efficiency can be further improved by miniaturizing the mechanical oscillator \cite{loncar}. In the context of solid-state defects in silicon carbide, coupling strengths can reach a few MHz \cite{SiC}, promising even more efficient cooling rates. Additionally, the use of magnons instead of NMOs could significantly enhance the spin-oscillator coupling, facilitating the experimental adoption of our protocol.

A notable challenge of our cooling approach is its reliance on post-selection, which results in a decreasing success rate as the number of probe projections increases. Nevertheless, our protocol offers a significant advantage for cooling large spin ensembles that are traditionally difficult to handle and cannot be cooled by standard optical or energy transfer techniques. For spin ensembles that are amenable to conventional cooling methods, such as NV centers in diamond, our protocol expands the cooling toolbox by working in conjunction with other methods, thus mitigating the problem of exponentially decreasing success rate. In addition, our analysis of a specific NV cantilever system suggests the possibility of an exciting experimental demonstration of partial ensemble polarization using currently available technologies.

This research highlights the significant impact that precise control of a single spin can have on the dynamics of a hybrid system, enabling energy transfer-like effects beyond the inherent capabilities of weak dispersive coupling. A similar idea has recently been demonstrated by an intriguing experiment: carefully designed inversion pulses on NV electronic spins can enable them to interact with extremely remote nuclear spins through paramagnetic spins (P1 centers) \cite{Goldblatt2024}.
In addition to remote sensing, our results also provide valuable insights into the preparation of complex quantum states in a macroscopic object, such as the GKP states.

\section*{Acknowledgement}

\noindent The authors would like to thank Jixing Zhang, Xin Zhang, and Ruoming Peng for their valuable feedback on the manuscript. This work was supported by the German Federal Ministry of Education and Research (BMBF) through the project QECHQS with Grant No. 16KIS1590K, and through the Future Cluster QSens. Additional support was provided by the Baden-Württemberg Stiftung through the project SPOC with Grant agreement No. QT-6 and through the EU Horizon Project SPINUS (101135699). The Deutsche Forschungsgemeinschaft (DFG) has funded this work under GRK 2642, and under Project No. 384846402 as part of FOR 2724.


\appendix

\section{Details of sketched experimental setup}

\label{sec:setup_details}

\noindent The probe spin is chosen as the electronic spin associated with a NV center due to its good spin coherence properties, high degree of spin control, and well resolved optical spectra at low temperatures \cite{Robledo2011,Abobeih2022,Bradley2019}. Its optical interface allows excitation to different optically excited energy levels that can either initialize or read out the spin state \cite{Aslam2017}. The NV electronic spins are also sensitive to their environment, allowing them to be (weakly) coupled to external spins, photons \cite{YangNatPhotonics2016,chikkaraddy2017} and phonons \cite{Browaeys2020,Peng2024}. 

The NV electronic spin is spin-1, which can be operated as a qubit with its $m_s=0, -1$ sublevels. This requires an external magnetic field  aligned with its symmetry axis to separate the $m_s = \pm 1$ levels, which is typically around 400 Gauss for NV centers in diamond \cite{Abobeih2022}. Such an arrangement negates the need to consider transverse spin-strain coupling due to its direct interaction between the $m_s = \pm 1$ levels \cite{Rabl2010NPhysics,Bennett2013PRL,Barfuss2015NatPhys}, allowing us to safely ignore the $m_s=+1$ sublevel in this work. What remains is the longitudinal strain coupling acting on the qubit levels, with a number that can be up to 10 kHz for NV centers \cite{lukinstrain}. This number can be even higher for other solid-state defect centers \cite{SiC}. 

In contrast, the spin ensemble to be cooled can be chosen more flexibly, since precise control of these spins is not a requirement for the current study. Potential targets include distant spins on other coupled cantilevers, spins at the clamped edge that are strain-coupled to the mechanical mode, or surface electron spins that are not well controlled. For example, if the ensemble spins are dark spins on the surface of the NMO \cite{HuilleryPRB2020} or other ferromagnetic particles \cite{Schomburg2011}, optical manipulation over them can be difficult, highlighting the need for other indirect polarization techniques.
Note that this ensemble-oscillator interaction generated by the attached magnetic tip can be 1–2 orders of magnitude stronger than the spin-strain coupling between the probe and the oscillator \cite{lukinprop}.

It is important to note that although we specifically consider a cantilever system, an alternative experimental realization could use the spin wave excitations of the spins (magnons) as the transducer \cite{Fukami2024NVMagnon,ViolaKusminskiy2019,Bejarano2024}. An advantage of this alternative is that the spin-magnon coupling can easily reach a few MHz \cite{Gonzalez2022PRB}, much stronger than the typical strength of spin-cantilever coupling by either strain or magnetic tip.

\section{Effects of inhomogeneous coupling and imperfect readout}

In this section, we briefly address the effects of other imperfections, such as inhomogeneous ensemble-oscillator coupling and imperfect probe readout. These factors can affect the cooling efficiency but do not impact the final achievable state.

\subsubsection{Inhomogeneous ensemble}
\noindent The inhomogeneity in coupling between the ensemble spins and the oscillator primarily arises from the spatial distribution of the gradient generated by the external magnetic field or the strain environment. Ref.~\cite{Rabl2009} demonstrates that for a spin ensemble located at a distance \( h = 10 \, \text{nm} \) from a magnetic tip and distributed over a planar area of \( 1 \, \mu \text{m}^2 \), the resulting inhomogeneity can be approximately \( g \pm g/10 \). As the gradient based decreases with distance from the tip as, $1/r$, spins spread over larger distances are effectively uncoupled to the oscillator. Minimizing the intra-dipolar couplings to be smaller than $g_0$ among the spins, $N \approx 10$ spins. Despite the presence of inhomogeneous couplings, our scheme still enables cooling of the spin ensemble, albeit at a reduced speed, necessitating more successful probe projections.

As illustrated in the upper panel of Fig.~\ref{fig:other_imperfections}, we sample an ensemble of four spins with randomly assigned coupling strengths to the oscillator, while keeping other parameters consistent with those in the main text. Compared to a uniformly coupled ensemble of the same size, both ensembles can be cooled to the ground state, although the inhomogeneous ensemble requires approximately 100 additional probe projections to achieve ground-state cooling. 

\subsubsection{Imperfect probe readout}
\noindent Imperfect readout of the probe spin (NV electronic spin considered in this work) often mainly occurs due to dark counts in the photodetector, which can mistakenly register incorrect projections as correct ones \cite{Bradley2019,Abobeih2022}. However, these erroneous measurements primarily affect the cooling rate, as they typically result in restarting the cooling process. 
Importantly, they do not impact the final steady state of the oscillator and the spin ensemble.

\begin{figure}
\vspace{-5mm}
\includegraphics[width=0.48\textwidth]{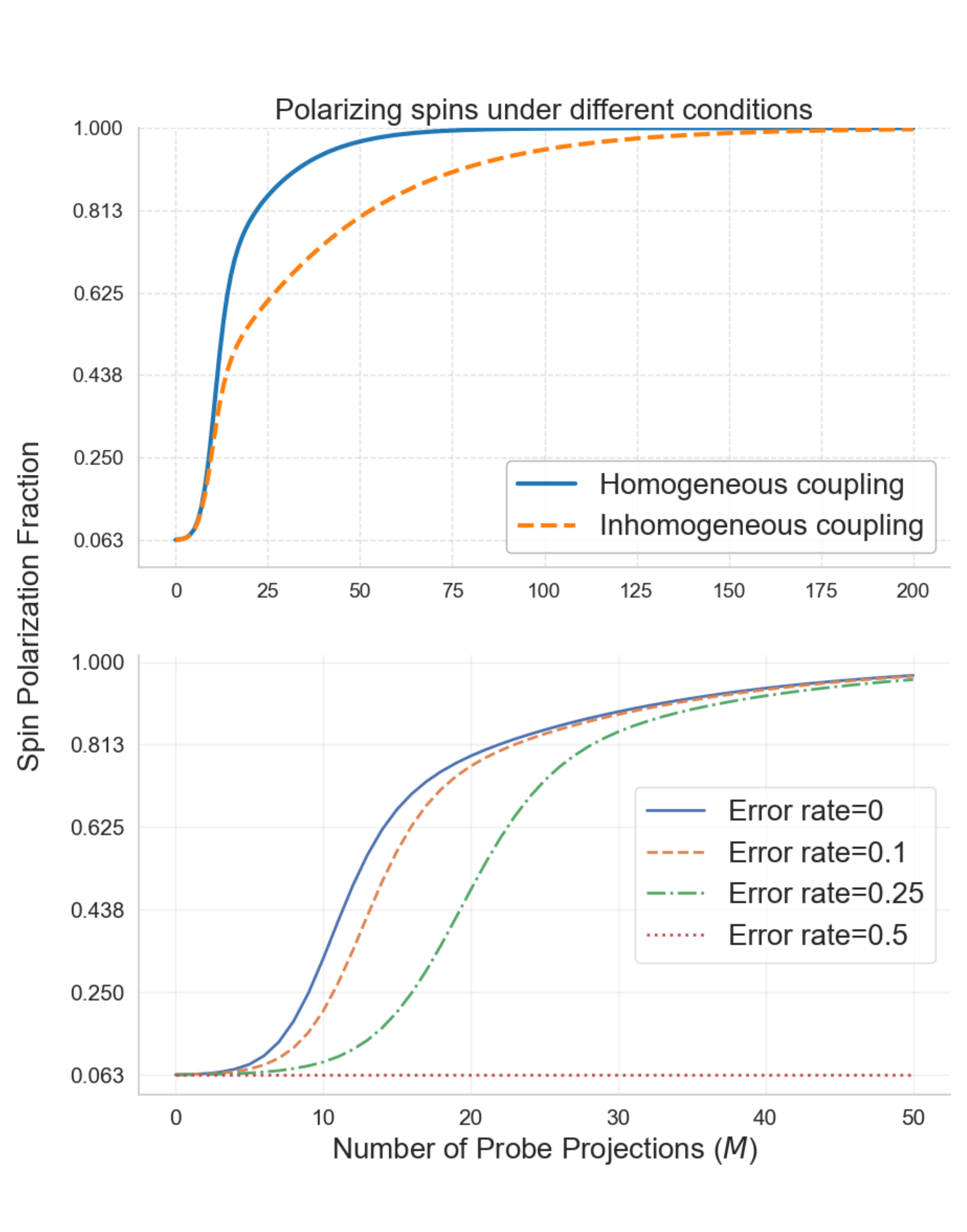}
\vspace{-10mm}
    \caption{
    Polarizing spin ensemble in the presence of inhomogeneous ensemble-oscillator coupling and imperfect probe readout.
    The parameters involved in the simulation are \(g/\omega = 0.01\), \(\epsilon/g = 0.15\), and \(n = 100\) pulses per probe projection.   
    (1) The upper panel shows the polarization of the spin ensemble reaching its maximum polarization sector, \(N/2\), for both homogeneous and inhomogeneous couplings of the spins to the oscillator, with the same mean value. 
    The inhomogeneous couplings are randomly sampled from a normal distribution with a mean value equal to $g$ and a standard deviation of $g/2$. In the homogeneous case, all four spins are identically coupled to the oscillator with strength \(g\).
    (2) The bottom panel illustrates the polarization of the spin ensemble reaching its maximum polarization sector, \(N/2\), under conditions of imperfect projective measurements of the probe. We consider a spin ensemble of four spins, each identically coupled to the oscillator. As the readout error increases, the cooling rate decreases, but the final steady state remains unaffected. In the extreme case where both projections are equally probable (error rate = 0.5), there is no cooling of the spin ensemble, confirming the validity of our measurement-based cooling scheme. 
    }
\label{fig:other_imperfections}
\end{figure}

In the bottom panel of Fig.~\ref{fig:other_imperfections}, we illustrate the polarization of the spin ensemble reaching its maximum polarization sector, \(N/2\), under conditions of imperfect projective measurements of the probe. We analyze a spin ensemble of four spins, each identically coupled to the oscillator. For the above simulations, the readout error is introduced as the probability that a projection to '-' is wrongly deemed as a successful probe projection into ‘+’ state. As the readout error rate increases, the cooling rate decreases, but the final steady state remains unaffected as shown in Fig.~\ref{fig:other_imperfections}. Even with an average readout error rate of 0.25, the ensemble can still be cooled to the ground state relatively quickly. Given that the readout error rate for NV centers can be as low as 0.05 \cite{Bradley2019,Abobeih2022}, we believe that readout error will not pose a significant bottleneck in experimentally realizing our cooling scheme.

It is important to note that in the extreme case where both projections are equally probable (readout error rate of 0.5), the ensemble spins keep unpolarized, thereby reinforcing the validity of our measurement-based cooling scheme.

\subsubsection{Effects of spin-$T_1$ and oscillator $Q$}
As the measurement based cooling scheme does not require coherences of the spin or the oscillator, the cooling fidelities are determined by their relative dissipation rates, $T_1$ for spins and $\gamma$ for the oscillator. In this work we have considered an ideal scenario i.e., infintie life time for the spins and the oscillator. Incorporating these effects will not be complex as no correlations are generated in the spin-mechanical system, and the probe measurements serve as an additional dissipation source for the oscillator and spin ensemble. The competition between the rates of their thermal equilibration and the measurement induced equilibration, results in a new steady state. 
For example in the presence of $T_1$, which tries to lower the maximal polarization $P^z_{N/2}$ towards $2/N$, at a rate $\gamma_{eq}$, and the measurement cooling increasing towards unity at a rate $\gamma_c$, the equilibrium polarization can be approximated using simple rate equations as
\begin{equation}
    P^z_{N/2}(M\rightarrow \infty) \sim \frac{\gamma_c}{\gamma_c + \gamma_{eq}}.
\end{equation}
For a typical range of $\gamma_c/\gamma_{eq} \ge 10$, the final polarization of the spin ensemble could always be larger than $90\%$.
The dissipative effects due to the oscillator have a dominant effect on the polarization fidelity. As the measurement-based cooling is based on the memory/correlation time of the oscillator state, a rapidly relaxing oscillator ($\gamma_{osc} \sim g_0$) will not lead to any effective cooling of the spins. 

\section{Simulation details}
\label{sec:simul_detail}

\noindent 
For the simulations conducted in the main text, the coupling parameters are also $g_0 = g = 10$ kHz, with an oscillator frequency of $\omega = 1.2$ MHz. Accordingly, we set the number of inversion pulses for each round of probe projection to $n=120$, see the probe control sequence in Fig.~\ref{eq:filter_in_limit}(b). This is to ensure that the effective probe-oscillator coupling is comparable to the oscillator frequency, as discussed in Eq.~(\ref{eq:filter_in_limit}).

In this work, we perform the simulation by calculating the dynamics directly on the initial Hamiltonian given in Eq.~(\ref{eq:initial_system_hamiltonian}) without any approximation. In addition, for simplicity, we approximate the oscillator mode as having only 100 levels, which is expected to be sufficient since we assume an initial oscillator thermal occupation of $n_0 \approx 45$ in the simulations.

\section{Encoding a qubit into a macroscopic oscillator}
\label{sec:GKPEncoding}

\noindent In this section, we consider a hybrid system consisting  only of the probe spin and the oscillator. We will show that we can prepare a macroscopic mechanical oscillator into a complicated quantum state using the same probe control sequence as our cooling protocol in the main text. The exploration of such quantum superposition states in macroscopic objects has emerged as a significant area of interest, mainly because of its potential to explore the frontier between classical and quantum physics \cite{Bild2023-ta,Fluhmann2019-jc}.

The quantum state in which we propose to prepare into a macroscopic mechanical oscillator is the so-called Gottesman-Kitaev-Preskill (GKP) state \cite{Gottesman2001-tl,Wang2017-im}, which is defined as the simultaneous eigenstates of the two stabilizers, i.e.,
\begin{equation}
    S_{q} = e^{-i2\sqrt{\pi}\hat{p}}, \, \, S_{p} = e^{i2\sqrt{\pi}\hat{q}}.
\end{equation}
The logical operators are correspondingly given by $X_{\scriptscriptstyle L} = \sqrt{S_{p}}$ and $Z_{\scriptscriptstyle L} = \sqrt{S_{q}}$, respectively. 

However, the ideal GKP states are unphysical, requiring infinite amount of energy to prepare \cite{Gottesman2001-tl}. Instead, we will consider the approximate GKP states that are defined as (up to normalization):
\begin{equation}
\begin{split}
&\ket{\tilde{0}} \propto \sum_{s=-\infty}^{\infty}\int_{-\infty}^\infty e^{-\frac{\Delta^2}{2}(2s)^2\pi}e^{-\frac{1}{2\Delta^2}(q-2s\sqrt{\pi})^2}\ket{q} \mathrm{d} q,  \\
&\ket{\tilde{1}} \propto \sum_{s=-\infty}^{\infty}\int_{-\infty}^\infty e^{-\frac{\Delta^2}{2}(2s+1)^2\pi}e^{-\frac{1}{2\Delta^2}(q-(2s+1)\sqrt{\pi})^2} \ket{q} \mathrm{d} q .
\end{split}
\label{eq:approximate_states}
\end{equation}
Note that the closer the squeezing parameter $\Delta$ is to 0, the higher is the quality of the GKP states \cite{Gottesman2001-tl,Wang2017-im}.

The GKP code is a promising candidate for quantum error correction (QEC) in continuous-variable systems \cite{Vuillot2019_qec}, such as trapped-ion mechanical oscillators \cite{Fluhmann2019-jc} and superconducting circuits \cite{Sivak2023}. In addition, GKP states hold the potential for enhanced sensing capabilities, allowing simultaneous detection of displacements in both the position and momentum quadratures of an oscillator \cite{Kasper2017_sensor}.

\subsection{Oscillator displacements along the momentum and position quadratures}

\noindent  
The projectors given in Eq.~(\(\ref{eq:oscillator_ensemble_projector}\)) now become just a displacement operator of the oscillator, which depends on the measurement result of the probe spin, as indicated by:
\begin{equation}
    \mathcal{P}_{\scriptscriptstyle N=0, \pm} = \frac{ D(\theta) \pm e^{i\phi}D(-\theta)}{2},
    \label{eq:projector_oscillator}
\end{equation}
where \(D(\theta) = \exp{(\theta a^{\dagger} - \theta^{*}a)}\) represents the displacement operator, with the amplitude \(\theta = g_0F(\epsilon, t)/2\omega\).

The additional phase \(\phi\) is introduced by rotating the probe spin by an angle \(\phi\) before the measurement, as shown in the circuit in Fig.~\ref{fig:GKP_phase_estimation_circuit}.
The proposed GKP encoding scheme involves repeated execution of this circuit, adaptively setting the angle \(\phi\) based on previous measurement results.  The angles \(\phi\) select the effective measurement basis of the probe spin. This is a phase estimation protocol that incrementally identifies the eigenvalue of $D(2\theta)$ while simultaneously projecting the oscillator into the corresponding eigenstate.

\subsubsection*{Displacement along position quadrature}
\noindent Our proposal is realized by carefully tuning the detuning \(\epsilon\) and applying a certain number of inversion pulses to the probe spin. For $\epsilon=0$, the filter function simplifies to $F(\epsilon, t) = 2(n+2)$ so that $\theta = g_0 t/\pi$. This leads to conditional displacements exactly along the position quadrature:
\begin{equation}
    D(\theta) = \exp\Big[\mp i\frac{\sqrt{2}g_0 t}{\pi}\hat{p}\Big], \,\, \text{with } \epsilon = 0,
\end{equation}
where $\hat{p} = \frac{i}{\sqrt{2}}(\hat{a} - \hat{a}^{\dagger})$ denotes the momentum operator of the oscillator. According to Ref.~\cite{Rosenfeld2021-ef}, this ability to adjust the oscillator's momentum based on the spin state facilitates a mechanism for entangling two spins by the oscillator.

\subsubsection*{Displacement along momentum quadrature}
\noindent Additionally, we also need to realize controlled oscillator-displacements along the momentum quadrature. This is achieved by setting $\epsilon = \omega/(n+1)$ so that $\epsilon t = \pi$, the filter function reads 
\begin{equation}
    F(\epsilon, t) = 2 + \frac{4}{e^{i\epsilon \tau}-1} = \frac{4}{i\epsilon\tau} + \mathcal{O}(\epsilon\tau),
\end{equation}
where $\hat{q} =  \frac{1}{\sqrt{2}}(\hat{a} + \hat{a}^{\dagger})$ is the position operator of the oscillator. Correspondingly, we have $\theta = ig_0 t(n+1)/\pi^2(n+2)$.
Furthermore, neglecting the higher order terms of $\epsilon\tau \sim \mathcal{O}(1/n)$ leads to the following displacement operator of the oscillator:
\begin{equation}
    D(\theta) = \exp\Big[\mp i\frac{2\sqrt{2}g_0 t}{\pi^2}\frac{n+1}{n+2}\hat{q}\Big], \, \, \text{with } \epsilon = \frac{\omega}{n+1}.
\label{eq:conditional_displacement_position}
\end{equation}

\subsubsection*{State preparation via phase estimation}

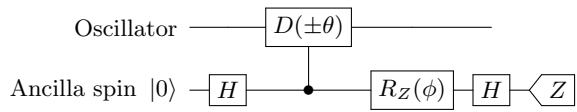
\begin{figure}
    \begin{minipage}{.5\textwidth}
        \Qcircuit @C=.8em @R=1em {
            \leftlabel{\mbox{Oscillator}} &\qw & \gate{D(\pm \theta)} & \qw & \qw \\
            \leftlabel{\mbox{Ancilla spin } \ket{0}} &\gate{H} & \ctrl{-1} &  \gate{R_Z(\phi)}  &\gate{H}  &\measuretab{Z}
        }
    \end{minipage}
    \caption{Single round of an adaptive phase estimation protocol that estimates the eigenvalue of oscillator displacement operator, $D(2\theta) = \exp{(2\theta a^{\dagger} - 2\theta^{*}a)}$ with amplitude $2\theta = g_0 F(\epsilon, t)/\omega$. Initially, the ancillary spin is set in the $\ket{+}$ state and subsequently measured in the x-basis after applying a series of periodic inversion pulses, which facilitates the implementation of controlled oscillator displacements. Before measuring the ancilla spin, it undergoes a z-axis rotation by an angle $\phi$. This angle $\phi$ is strategically determined based on the ancilla measurement results of previous phase estimation cycles, with the goal of minimize the uncertainty in the posterior probability distribution estimate of $D(2\theta)$'s eigenvalue. Specifically, when $\phi=0$, this circuit simply executes the projectors given in Eq.~(\ref{eq:projector_oscillator}).  Through repeated application of this phase estimation circuit, the oscillator states outside the subspace spanned by the eigenstates of \( D(2\theta) \) are gradually suppressed.}
\label{fig:GKP_phase_estimation_circuit}
\end{figure}

\noindent 
Phase estimation has been shown to be effective for preparing such approximate GKP logical states by determining the eigenvalues of stabilizers and logical operators \cite{Terhal2016-sy}. For the spin-oscillator hybrid system considered here, such eigenvalue determination is realized by adjusting the detuning magnitude $\epsilon$ and the total evolution duration $t$. In this way, we can choose the magnitude by which the oscillator is shifted along the momentum or position quadrature.

For example, to prepare the approximate GKP state \( \ket{\tilde{0}} \), first initialize the oscillator in its ground state with zero photons. It then sets \( D(\theta) = S_{q} \) and runs the phase estimation circuit several times to estimate its eigenvalue with high confidence. A subsequent fit to \( D(\theta) = Z_{\scriptscriptstyle L} \) allows for similar eigenvalue estimation. Determining the eigenvalues of $S_{q}$ and $Z_{\scriptscriptstyle L}$ projects the oscillator into the desired approximate GKP state. This process can be described as iterative and adaptive application of the circuit in Fig.~\ref{fig:GKP_phase_estimation_circuit}.

\subsubsection*{GKP encoding and oscillator cooling}
\noindent The GKP encoding scheme we discuss is similar to the oscillator cooling technique described in Ref.~\cite{Rao2016-uw}, with both methods using the phase estimation circuit shown in Fig.~\ref{fig:GKP_phase_estimation_circuit} many times. However, the encoding scheme starts from the ground state, as opposed to the cooling method, which applies the circuit to a thermal state and then cools it down to the ground state.

A crucial difference between the two lies in how the circuit is used in the cooling process, where it is implemented with non-zero detuning, so that neither Eq.(16) nor Eq.(18) hold. When these conditions are satisfied, instead of cooling to a zero photon ground state, the oscillator, initially in a thermal state, stabilizes in a state that maintains a finite number of photons \cite{Rao2016-uw}.

\subsection{Experimental feasibility}
\noindent 
In contrast to the oscillator cooling scheme developed in Ref.~\cite{Rao2016-uw}, which requires only a weak probe-oscillator coupling, the GKP encoding scheme is much more demanding as it requires a strong coupling strength. In this respect, current experimental capabilities are clearly insufficient to implement the preparation scheme. 
Moreover, the GKP encoding scheme has been extensively studied in Ref.~\cite{Terhal2016-sy}. 
Therefore, we choose not to perform detailed numerical simulations to evaluate the practicality of the GKP encoding scheme in this study.
Instead, we briefly assess the challenges and discuss potential improvements to existing setups that could facilitate the preparation of GKP states in macroscopic mechanical oscillators.

The time duration for each phase estimation round needs to be of the order of \( 1/g_0 \), so that the displacement magnitude can be of the order of \( |\theta| \sim O(\sqrt{\pi}) \). Given the necessity of around 10 phase estimation cycles to yield an approximate GKP state of acceptable quality \cite{Terhal2016-sy}, the preparation scheme requires an overall time exceeding \( \mathcal{O}(10/g_0) \).

From an experimental point of view, it's crucial that the coherence time of the oscillator is considerably long compared to the coupling strength $g_0$, ideally at least on the order of $100/g_0$. Diamond NMOs with a quality factor $Q > 10^6$ have been demonstrated at sub-Kelvin temperatures and mechanical damping rates of a few kHz \cite{degen}. This implies that the coherence time of the vibrational mode must be significantly improved. Note that the coherence time of the NV electronic spin in cryogenic conditions can exceed 1 second by the application of dynamical decoupling \cite{Abobeih2018}. This is more than sufficient, as the probe spin only needs to maintain coherence during each phase estimation round.

\bibliography{main.bib}

\begin{thebibliography}{58}%
\makeatletter
\providecommand \@ifxundefined [1]{%
 \@ifx{#1\undefined}
}%
\providecommand \@ifnum [1]{%
 \ifnum #1\expandafter \@firstoftwo
 \else \expandafter \@secondoftwo
 \fi
}%
\providecommand \@ifx [1]{%
 \ifx #1\expandafter \@firstoftwo
 \else \expandafter \@secondoftwo
 \fi
}%
\providecommand \natexlab [1]{#1}%
\providecommand \enquote  [1]{``#1''}%
\providecommand \bibnamefont  [1]{#1}%
\providecommand \bibfnamefont [1]{#1}%
\providecommand \citenamefont [1]{#1}%
\providecommand \href@noop [0]{\@secondoftwo}%
\providecommand \href [0]{\begingroup \@sanitize@url \@href}%
\providecommand \@href[1]{\@@startlink{#1}\@@href}%
\providecommand \@@href[1]{\endgroup#1\@@endlink}%
\providecommand \@sanitize@url [0]{\catcode `\\12\catcode `\$12\catcode `\&12\catcode `\#12\catcode `\^12\catcode `\_12\catcode `\%12\relax}%
\providecommand \@@startlink[1]{}%
\providecommand \@@endlink[0]{}%
\providecommand \url  [0]{\begingroup\@sanitize@url \@url }%
\providecommand \@url [1]{\endgroup\@href {#1}{\urlprefix }}%
\providecommand \urlprefix  [0]{URL }%
\providecommand \Eprint [0]{\href }%
\providecommand \doibase [0]{https://doi.org/}%
\providecommand \selectlanguage [0]{\@gobble}%
\providecommand \bibinfo  [0]{\@secondoftwo}%
\providecommand \bibfield  [0]{\@secondoftwo}%
\providecommand \translation [1]{[#1]}%
\providecommand \BibitemOpen [0]{}%
\providecommand \bibitemStop [0]{}%
\providecommand \bibitemNoStop [0]{.\EOS\space}%
\providecommand \EOS [0]{\spacefactor3000\relax}%
\providecommand \BibitemShut  [1]{\csname bibitem#1\endcsname}%
\let\auto@bib@innerbib\@empty
\bibitem [{\citenamefont {Battistel}\ \emph {et~al.}(2023)\citenamefont {Battistel}, \citenamefont {Chamberland}, \citenamefont {Johar}, \citenamefont {Overwater}, \citenamefont {Sebastiano}, \citenamefont {Skoric}, \citenamefont {Ueno},\ and\ \citenamefont {Usman}}]{Battistel2023Challenges}%
  \BibitemOpen
  \bibfield  {author} {\bibinfo {author} {\bibfnamefont {F.}~\bibnamefont {Battistel}}, \bibinfo {author} {\bibfnamefont {C.}~\bibnamefont {Chamberland}}, \bibinfo {author} {\bibfnamefont {K.}~\bibnamefont {Johar}}, \bibinfo {author} {\bibfnamefont {R.~W.~J.}\ \bibnamefont {Overwater}}, \bibinfo {author} {\bibfnamefont {F.}~\bibnamefont {Sebastiano}}, \bibinfo {author} {\bibfnamefont {L.}~\bibnamefont {Skoric}}, \bibinfo {author} {\bibfnamefont {Y.}~\bibnamefont {Ueno}},\ and\ \bibinfo {author} {\bibfnamefont {M.}~\bibnamefont {Usman}},\ }\bibfield  {title} {\bibinfo {title} {Real-time decoding for fault-tolerant quantum computing: progress, challenges and outlook},\ }\href {https://doi.org/10.1088/2399-1984/aceba6} {\bibfield  {journal} {\bibinfo  {journal} {Nano Futures}\ }\textbf {\bibinfo {volume} {7}},\ \bibinfo {pages} {032003} (\bibinfo {year} {2023})}\BibitemShut {NoStop}%
\bibitem [{\citenamefont {Lordi}\ and\ \citenamefont {Nichol}(2021)}]{Lordi2021}%
  \BibitemOpen
  \bibfield  {author} {\bibinfo {author} {\bibfnamefont {V.}~\bibnamefont {Lordi}}\ and\ \bibinfo {author} {\bibfnamefont {J.~M.}\ \bibnamefont {Nichol}},\ }\bibfield  {title} {\bibinfo {title} {Advances and opportunities in materials science for scalable quantum computing},\ }\href {https://doi.org/10.1557/s43577-021-00133-0} {\bibfield  {journal} {\bibinfo  {journal} {MRS Bulletin}\ }\textbf {\bibinfo {volume} {46}},\ \bibinfo {pages} {589} (\bibinfo {year} {2021})}\BibitemShut {NoStop}%
\bibitem [{\citenamefont {Xiang}\ \emph {et~al.}(2013)\citenamefont {Xiang}, \citenamefont {Ashhab}, \citenamefont {You},\ and\ \citenamefont {Nori}}]{XiangRMP2013}%
  \BibitemOpen
  \bibfield  {author} {\bibinfo {author} {\bibfnamefont {Z.-L.}\ \bibnamefont {Xiang}}, \bibinfo {author} {\bibfnamefont {S.}~\bibnamefont {Ashhab}}, \bibinfo {author} {\bibfnamefont {J.~Q.}\ \bibnamefont {You}},\ and\ \bibinfo {author} {\bibfnamefont {F.}~\bibnamefont {Nori}},\ }\bibfield  {title} {\bibinfo {title} {Hybrid quantum circuits: Superconducting circuits interacting with other quantum systems},\ }\href {https://doi.org/10.1103/RevModPhys.85.623} {\bibfield  {journal} {\bibinfo  {journal} {Rev. Mod. Phys.}\ }\textbf {\bibinfo {volume} {85}},\ \bibinfo {pages} {623} (\bibinfo {year} {2013})}\BibitemShut {NoStop}%
\bibitem [{\citenamefont {Kurizki}\ \emph {et~al.}(2015)\citenamefont {Kurizki}, \citenamefont {Bertet}, \citenamefont {Kubo}, \citenamefont {Mølmer}, \citenamefont {Petrosyan}, \citenamefont {Rabl},\ and\ \citenamefont {Schmiedmayer}}]{Kurizki2015}%
  \BibitemOpen
  \bibfield  {author} {\bibinfo {author} {\bibfnamefont {G.}~\bibnamefont {Kurizki}}, \bibinfo {author} {\bibfnamefont {P.}~\bibnamefont {Bertet}}, \bibinfo {author} {\bibfnamefont {Y.}~\bibnamefont {Kubo}}, \bibinfo {author} {\bibfnamefont {K.}~\bibnamefont {Mølmer}}, \bibinfo {author} {\bibfnamefont {D.}~\bibnamefont {Petrosyan}}, \bibinfo {author} {\bibfnamefont {P.}~\bibnamefont {Rabl}},\ and\ \bibinfo {author} {\bibfnamefont {J.}~\bibnamefont {Schmiedmayer}},\ }\bibfield  {title} {\bibinfo {title} {Quantum technologies with hybrid systems},\ }\href {https://doi.org/10.1073/pnas.1419326112} {\bibfield  {journal} {\bibinfo  {journal} {Proc. Natl. Acad. Sci.}\ }\textbf {\bibinfo {volume} {112}},\ \bibinfo {pages} {3866} (\bibinfo {year} {2015})}\BibitemShut {NoStop}%
\bibitem [{\citenamefont {Hei}\ \emph {et~al.}(2023)\citenamefont {Hei}, \citenamefont {Li}, \citenamefont {Pan},\ and\ \citenamefont {Nori}}]{HeiPRL2023}%
  \BibitemOpen
  \bibfield  {author} {\bibinfo {author} {\bibfnamefont {X.-L.}\ \bibnamefont {Hei}}, \bibinfo {author} {\bibfnamefont {P.-B.}\ \bibnamefont {Li}}, \bibinfo {author} {\bibfnamefont {X.-F.}\ \bibnamefont {Pan}},\ and\ \bibinfo {author} {\bibfnamefont {F.}~\bibnamefont {Nori}},\ }\bibfield  {title} {\bibinfo {title} {Enhanced tripartite interactions in spin-magnon-mechanical hybrid systems},\ }\href {https://doi.org/10.1103/PhysRevLett.130.073602} {\bibfield  {journal} {\bibinfo  {journal} {Phys. Rev. Lett.}\ }\textbf {\bibinfo {volume} {130}},\ \bibinfo {pages} {073602} (\bibinfo {year} {2023})}\BibitemShut {NoStop}%
\bibitem [{\citenamefont {Kolkowitz}\ \emph {et~al.}(2012)\citenamefont {Kolkowitz}, \citenamefont {Bleszynski~Jayich}, \citenamefont {Unterreithmeier}, \citenamefont {Bennett}, \citenamefont {Rabl}, \citenamefont {Harris},\ and\ \citenamefont {Lukin}}]{Kolkowitz2012Sensing}%
  \BibitemOpen
  \bibfield  {author} {\bibinfo {author} {\bibfnamefont {S.}~\bibnamefont {Kolkowitz}}, \bibinfo {author} {\bibfnamefont {A.~C.}\ \bibnamefont {Bleszynski~Jayich}}, \bibinfo {author} {\bibfnamefont {Q.~P.}\ \bibnamefont {Unterreithmeier}}, \bibinfo {author} {\bibfnamefont {S.~D.}\ \bibnamefont {Bennett}}, \bibinfo {author} {\bibfnamefont {P.}~\bibnamefont {Rabl}}, \bibinfo {author} {\bibfnamefont {J.~G.~E.}\ \bibnamefont {Harris}},\ and\ \bibinfo {author} {\bibfnamefont {M.~D.}\ \bibnamefont {Lukin}},\ }\bibfield  {title} {\bibinfo {title} {Coherent sensing of a mechanical resonator with a single-spin qubit},\ }\href {https://doi.org/10.1126/science.1216821} {\bibfield  {journal} {\bibinfo  {journal} {Science}\ }\textbf {\bibinfo {volume} {335}},\ \bibinfo {pages} {1603} (\bibinfo {year} {2012})}\BibitemShut {NoStop}%
\bibitem [{\citenamefont {Rabl}\ \emph {et~al.}(2010)\citenamefont {Rabl}, \citenamefont {Kolkowitz}, \citenamefont {Koppens}, \citenamefont {Harris}, \citenamefont {Zoller},\ and\ \citenamefont {Lukin}}]{Rabl2010NPhysics}%
  \BibitemOpen
  \bibfield  {author} {\bibinfo {author} {\bibfnamefont {P.}~\bibnamefont {Rabl}}, \bibinfo {author} {\bibfnamefont {S.~J.}\ \bibnamefont {Kolkowitz}}, \bibinfo {author} {\bibfnamefont {F.~H.~L.}\ \bibnamefont {Koppens}}, \bibinfo {author} {\bibfnamefont {J.~G.~E.}\ \bibnamefont {Harris}}, \bibinfo {author} {\bibfnamefont {P.}~\bibnamefont {Zoller}},\ and\ \bibinfo {author} {\bibfnamefont {M.~D.}\ \bibnamefont {Lukin}},\ }\bibfield  {title} {\bibinfo {title} {A quantum spin transducer based on nanoelectromechanical resonator arrays},\ }\href {https://doi.org/10.1038/nphys1679} {\bibfield  {journal} {\bibinfo  {journal} {Nat. Phys.}\ }\textbf {\bibinfo {volume} {6}},\ \bibinfo {pages} {602} (\bibinfo {year} {2010})}\BibitemShut {NoStop}%
\bibitem [{\citenamefont {Yang}\ \emph {et~al.}(2016)\citenamefont {Yang}, \citenamefont {Wang}, \citenamefont {Rao}, \citenamefont {Hien~Tran}, \citenamefont {Momenzadeh}, \citenamefont {Markham}, \citenamefont {Twitchen}, \citenamefont {Wang}, \citenamefont {Yang}, \citenamefont {Stöhr}, \citenamefont {Neumann}, \citenamefont {Kosaka},\ and\ \citenamefont {Wrachtrup}}]{YangNatPhotonics2016}%
  \BibitemOpen
  \bibfield  {author} {\bibinfo {author} {\bibfnamefont {S.}~\bibnamefont {Yang}}, \bibinfo {author} {\bibfnamefont {Y.}~\bibnamefont {Wang}}, \bibinfo {author} {\bibfnamefont {D.~D.~B.}\ \bibnamefont {Rao}}, \bibinfo {author} {\bibfnamefont {T.}~\bibnamefont {Hien~Tran}}, \bibinfo {author} {\bibfnamefont {A.~S.}\ \bibnamefont {Momenzadeh}}, \bibinfo {author} {\bibfnamefont {M.}~\bibnamefont {Markham}}, \bibinfo {author} {\bibfnamefont {D.~J.}\ \bibnamefont {Twitchen}}, \bibinfo {author} {\bibfnamefont {P.}~\bibnamefont {Wang}}, \bibinfo {author} {\bibfnamefont {W.}~\bibnamefont {Yang}}, \bibinfo {author} {\bibfnamefont {R.}~\bibnamefont {Stöhr}}, \bibinfo {author} {\bibfnamefont {P.}~\bibnamefont {Neumann}}, \bibinfo {author} {\bibfnamefont {H.}~\bibnamefont {Kosaka}},\ and\ \bibinfo {author} {\bibfnamefont {J.}~\bibnamefont {Wrachtrup}},\ }\bibfield  {title} {\bibinfo {title} {High-fidelity transfer and storage of photon states in a single nuclear spin},\ }\href {https://doi.org/10.1038/nphoton.2016.103}
  {\bibfield  {journal} {\bibinfo  {journal} {Nat. Photonics}\ }\textbf {\bibinfo {volume} {10}},\ \bibinfo {pages} {507} (\bibinfo {year} {2016})}\BibitemShut {NoStop}%
\bibitem [{\citenamefont {Browaeys}\ and\ \citenamefont {Lahaye}(2020)}]{Browaeys2020}%
  \BibitemOpen
  \bibfield  {author} {\bibinfo {author} {\bibfnamefont {A.}~\bibnamefont {Browaeys}}\ and\ \bibinfo {author} {\bibfnamefont {T.}~\bibnamefont {Lahaye}},\ }\bibfield  {title} {\bibinfo {title} {Many-body physics with individually controlled rydberg atoms},\ }\href {https://doi.org/10.1038/s41567-019-0733-z} {\bibfield  {journal} {\bibinfo  {journal} {Nat. Phys.}\ }\textbf {\bibinfo {volume} {16}},\ \bibinfo {pages} {132} (\bibinfo {year} {2020})}\BibitemShut {NoStop}%
\bibitem [{\citenamefont {Viola~Kusminskiy}(2019)}]{ViolaKusminskiy2019}%
  \BibitemOpen
  \bibfield  {author} {\bibinfo {author} {\bibfnamefont {S.}~\bibnamefont {Viola~Kusminskiy}},\ }\href {https://doi.org/10.1007/978-3-030-13345-0} {\emph {\bibinfo {title} {Quantum Magnetism, Spin Waves, and Optical Cavities}}}\ (\bibinfo  {publisher} {Springer International Publishing},\ \bibinfo {year} {2019})\BibitemShut {NoStop}%
\bibitem [{\citenamefont {Bejarano}\ \emph {et~al.}(2024)\citenamefont {Bejarano}, \citenamefont {Goncalves}, \citenamefont {Hache}, \citenamefont {Hollenbach}, \citenamefont {Heins}, \citenamefont {Hula}, \citenamefont {Körber}, \citenamefont {Heinze}, \citenamefont {Berencén}, \citenamefont {Helm}, \citenamefont {Fassbender}, \citenamefont {Astakhov},\ and\ \citenamefont {Schultheiss}}]{Bejarano2024}%
  \BibitemOpen
  \bibfield  {author} {\bibinfo {author} {\bibfnamefont {M.}~\bibnamefont {Bejarano}}, \bibinfo {author} {\bibfnamefont {F.~J.~T.}\ \bibnamefont {Goncalves}}, \bibinfo {author} {\bibfnamefont {T.}~\bibnamefont {Hache}}, \bibinfo {author} {\bibfnamefont {M.}~\bibnamefont {Hollenbach}}, \bibinfo {author} {\bibfnamefont {C.}~\bibnamefont {Heins}}, \bibinfo {author} {\bibfnamefont {T.}~\bibnamefont {Hula}}, \bibinfo {author} {\bibfnamefont {L.}~\bibnamefont {Körber}}, \bibinfo {author} {\bibfnamefont {J.}~\bibnamefont {Heinze}}, \bibinfo {author} {\bibfnamefont {Y.}~\bibnamefont {Berencén}}, \bibinfo {author} {\bibfnamefont {M.}~\bibnamefont {Helm}}, \bibinfo {author} {\bibfnamefont {J.}~\bibnamefont {Fassbender}}, \bibinfo {author} {\bibfnamefont {G.~V.}\ \bibnamefont {Astakhov}},\ and\ \bibinfo {author} {\bibfnamefont {H.}~\bibnamefont {Schultheiss}},\ }\bibfield  {title} {\bibinfo {title} {Parametric magnon transduction to spin qubits},\ }\bibfield  {journal} {\bibinfo  {journal} {Sci. Adv.}\ }\textbf
  {\bibinfo {volume} {10}},\ \href {https://doi.org/10.1126/sciadv.adi2042} {10.1126/sciadv.adi2042} (\bibinfo {year} {2024})\BibitemShut {NoStop}%
\bibitem [{\citenamefont {Staudacher}\ \emph {et~al.}(2013)\citenamefont {Staudacher}, \citenamefont {Shi}, \citenamefont {Pezzagna}, \citenamefont {Meijer}, \citenamefont {Du}, \citenamefont {Meriles}, \citenamefont {Reinhard},\ and\ \citenamefont {Wrachtrup}}]{Staudacher2013}%
  \BibitemOpen
  \bibfield  {author} {\bibinfo {author} {\bibfnamefont {T.}~\bibnamefont {Staudacher}}, \bibinfo {author} {\bibfnamefont {F.}~\bibnamefont {Shi}}, \bibinfo {author} {\bibfnamefont {S.}~\bibnamefont {Pezzagna}}, \bibinfo {author} {\bibfnamefont {J.}~\bibnamefont {Meijer}}, \bibinfo {author} {\bibfnamefont {J.}~\bibnamefont {Du}}, \bibinfo {author} {\bibfnamefont {C.~A.}\ \bibnamefont {Meriles}}, \bibinfo {author} {\bibfnamefont {F.}~\bibnamefont {Reinhard}},\ and\ \bibinfo {author} {\bibfnamefont {J.}~\bibnamefont {Wrachtrup}},\ }\bibfield  {title} {\bibinfo {title} {Nuclear magnetic resonance spectroscopy on a (5-nanometer) 3 sample volume},\ }\href {https://doi.org/10.1126/science.1231675} {\bibfield  {journal} {\bibinfo  {journal} {Science}\ }\textbf {\bibinfo {volume} {339}},\ \bibinfo {pages} {561} (\bibinfo {year} {2013})}\BibitemShut {NoStop}%
\bibitem [{\citenamefont {Goldblatt}\ \emph {et~al.}(2024)\citenamefont {Goldblatt}, \citenamefont {Martin},\ and\ \citenamefont {Wood}}]{Goldblatt2024}%
  \BibitemOpen
  \bibfield  {author} {\bibinfo {author} {\bibfnamefont {R.}~\bibnamefont {Goldblatt}}, \bibinfo {author} {\bibfnamefont {A.}~\bibnamefont {Martin}},\ and\ \bibinfo {author} {\bibfnamefont {A.}~\bibnamefont {Wood}},\ }\bibfield  {title} {\bibinfo {title} {Sensing coherent nuclear spin dynamics with an ensemble of paramagnetic nitrogen spins},\ }\href {https://doi.org/10.1103/PRXQuantum.5.020334} {\bibfield  {journal} {\bibinfo  {journal} {PRX Quantum}\ }\textbf {\bibinfo {volume} {5}},\ \bibinfo {pages} {020334} (\bibinfo {year} {2024})}\BibitemShut {NoStop}%
\bibitem [{\citenamefont {van~de Stolpe}\ \emph {et~al.}(2024)\citenamefont {van~de Stolpe}, \citenamefont {Kwiatkowski}, \citenamefont {Bradley}, \citenamefont {Randall}, \citenamefont {Abobeih}, \citenamefont {Breitweiser}, \citenamefont {Bassett}, \citenamefont {Markham}, \citenamefont {Twitchen},\ and\ \citenamefont {Taminiau}}]{Stolpe2024}%
  \BibitemOpen
  \bibfield  {author} {\bibinfo {author} {\bibfnamefont {G.~L.}\ \bibnamefont {van~de Stolpe}}, \bibinfo {author} {\bibfnamefont {D.~P.}\ \bibnamefont {Kwiatkowski}}, \bibinfo {author} {\bibfnamefont {C.~E.}\ \bibnamefont {Bradley}}, \bibinfo {author} {\bibfnamefont {J.}~\bibnamefont {Randall}}, \bibinfo {author} {\bibfnamefont {M.~H.}\ \bibnamefont {Abobeih}}, \bibinfo {author} {\bibfnamefont {S.~A.}\ \bibnamefont {Breitweiser}}, \bibinfo {author} {\bibfnamefont {L.~C.}\ \bibnamefont {Bassett}}, \bibinfo {author} {\bibfnamefont {M.}~\bibnamefont {Markham}}, \bibinfo {author} {\bibfnamefont {D.~J.}\ \bibnamefont {Twitchen}},\ and\ \bibinfo {author} {\bibfnamefont {T.~H.}\ \bibnamefont {Taminiau}},\ }\bibfield  {title} {\bibinfo {title} {Mapping a 50-spin-qubit network through correlated sensing},\ }\bibfield  {journal} {\bibinfo  {journal} {Nat. Commun.}\ }\textbf {\bibinfo {volume} {15}},\ \href {https://doi.org/10.1038/s41467-024-46075-4} {10.1038/s41467-024-46075-4} (\bibinfo {year} {2024})\BibitemShut
  {NoStop}%
\bibitem [{\citenamefont {Abobeih}\ \emph {et~al.}(2019)\citenamefont {Abobeih}, \citenamefont {Randall}, \citenamefont {Bradley}, \citenamefont {Bartling}, \citenamefont {Bakker}, \citenamefont {Degen}, \citenamefont {Markham}, \citenamefont {Twitchen},\ and\ \citenamefont {Taminiau}}]{Abobeih2019}%
  \BibitemOpen
  \bibfield  {author} {\bibinfo {author} {\bibfnamefont {M.~H.}\ \bibnamefont {Abobeih}}, \bibinfo {author} {\bibfnamefont {J.}~\bibnamefont {Randall}}, \bibinfo {author} {\bibfnamefont {C.~E.}\ \bibnamefont {Bradley}}, \bibinfo {author} {\bibfnamefont {H.~P.}\ \bibnamefont {Bartling}}, \bibinfo {author} {\bibfnamefont {M.~A.}\ \bibnamefont {Bakker}}, \bibinfo {author} {\bibfnamefont {M.~J.}\ \bibnamefont {Degen}}, \bibinfo {author} {\bibfnamefont {M.}~\bibnamefont {Markham}}, \bibinfo {author} {\bibfnamefont {D.~J.}\ \bibnamefont {Twitchen}},\ and\ \bibinfo {author} {\bibfnamefont {T.~H.}\ \bibnamefont {Taminiau}},\ }\bibfield  {title} {\bibinfo {title} {{Atomic-scale imaging of a 27-nuclear-spin cluster using a quantum sensor}},\ }\href {https://doi.org/10.1038/s41586-019-1834-7} {\bibfield  {journal} {\bibinfo  {journal} {Nature}\ }\textbf {\bibinfo {volume} {576}},\ \bibinfo {pages} {411} (\bibinfo {year} {2019})}\BibitemShut {NoStop}%
\bibitem [{\citenamefont {Schlipf}\ \emph {et~al.}(2017)\citenamefont {Schlipf}, \citenamefont {Oeckinghaus}, \citenamefont {Xu}, \citenamefont {Dasari}, \citenamefont {Zappe}, \citenamefont {de~Oliveira}, \citenamefont {Kern}, \citenamefont {Azarkh}, \citenamefont {Drescher}, \citenamefont {Ternes}, \citenamefont {Kern}, \citenamefont {Wrachtrup},\ and\ \citenamefont {Finkler}}]{Schlipf2017}%
  \BibitemOpen
  \bibfield  {author} {\bibinfo {author} {\bibfnamefont {L.}~\bibnamefont {Schlipf}}, \bibinfo {author} {\bibfnamefont {T.}~\bibnamefont {Oeckinghaus}}, \bibinfo {author} {\bibfnamefont {K.}~\bibnamefont {Xu}}, \bibinfo {author} {\bibfnamefont {D.~B.~R.}\ \bibnamefont {Dasari}}, \bibinfo {author} {\bibfnamefont {A.}~\bibnamefont {Zappe}}, \bibinfo {author} {\bibfnamefont {F.~F.}\ \bibnamefont {de~Oliveira}}, \bibinfo {author} {\bibfnamefont {B.}~\bibnamefont {Kern}}, \bibinfo {author} {\bibfnamefont {M.}~\bibnamefont {Azarkh}}, \bibinfo {author} {\bibfnamefont {M.}~\bibnamefont {Drescher}}, \bibinfo {author} {\bibfnamefont {M.}~\bibnamefont {Ternes}}, \bibinfo {author} {\bibfnamefont {K.}~\bibnamefont {Kern}}, \bibinfo {author} {\bibfnamefont {J.}~\bibnamefont {Wrachtrup}},\ and\ \bibinfo {author} {\bibfnamefont {A.}~\bibnamefont {Finkler}},\ }\bibfield  {title} {\bibinfo {title} {A molecular quantum spin network controlled by a single qubit},\ }\bibfield  {journal} {\bibinfo  {journal} {Science Advances}\
  }\textbf {\bibinfo {volume} {3}},\ \href {https://doi.org/10.1126/sciadv.1701116} {10.1126/sciadv.1701116} (\bibinfo {year} {2017})\BibitemShut {NoStop}%
\bibitem [{\citenamefont {Fukami}\ \emph {et~al.}(2024)\citenamefont {Fukami}, \citenamefont {Marcks}, \citenamefont {Candido}, \citenamefont {Weiss}, \citenamefont {Soloway}, \citenamefont {Sullivan}, \citenamefont {Delegan}, \citenamefont {Heremans}, \citenamefont {Flatté},\ and\ \citenamefont {Awschalom}}]{Fukami2024NVMagnon}%
  \BibitemOpen
  \bibfield  {author} {\bibinfo {author} {\bibfnamefont {M.}~\bibnamefont {Fukami}}, \bibinfo {author} {\bibfnamefont {J.~C.}\ \bibnamefont {Marcks}}, \bibinfo {author} {\bibfnamefont {D.~R.}\ \bibnamefont {Candido}}, \bibinfo {author} {\bibfnamefont {L.~R.}\ \bibnamefont {Weiss}}, \bibinfo {author} {\bibfnamefont {B.}~\bibnamefont {Soloway}}, \bibinfo {author} {\bibfnamefont {S.~E.}\ \bibnamefont {Sullivan}}, \bibinfo {author} {\bibfnamefont {N.}~\bibnamefont {Delegan}}, \bibinfo {author} {\bibfnamefont {F.~J.}\ \bibnamefont {Heremans}}, \bibinfo {author} {\bibfnamefont {M.~E.}\ \bibnamefont {Flatté}},\ and\ \bibinfo {author} {\bibfnamefont {D.~D.}\ \bibnamefont {Awschalom}},\ }\bibfield  {title} {\bibinfo {title} {Magnon-mediated qubit coupling determined via dissipation measurements},\ }\bibfield  {journal} {\bibinfo  {journal} {Proc. Natl. Acad. Sci.}\ }\textbf {\bibinfo {volume} {121}},\ \href {https://doi.org/10.1073/pnas.2313754120} {10.1073/pnas.2313754120} (\bibinfo {year} {2024})\BibitemShut
  {NoStop}%
\bibitem [{\citenamefont {Gonzalez-Ballestero}\ \emph {et~al.}(2020)\citenamefont {Gonzalez-Ballestero}, \citenamefont {H\"ummer}, \citenamefont {Gieseler},\ and\ \citenamefont {Romero-Isart}}]{prbrefisart}%
  \BibitemOpen
  \bibfield  {author} {\bibinfo {author} {\bibfnamefont {C.}~\bibnamefont {Gonzalez-Ballestero}}, \bibinfo {author} {\bibfnamefont {D.}~\bibnamefont {H\"ummer}}, \bibinfo {author} {\bibfnamefont {J.}~\bibnamefont {Gieseler}},\ and\ \bibinfo {author} {\bibfnamefont {O.}~\bibnamefont {Romero-Isart}},\ }\bibfield  {title} {\bibinfo {title} {Theory of quantum acoustomagnonics and acoustomechanics with a micromagnet},\ }\href {https://doi.org/10.1103/PhysRevB.101.125404} {\bibfield  {journal} {\bibinfo  {journal} {Phys. Rev. B}\ }\textbf {\bibinfo {volume} {101}},\ \bibinfo {pages} {125404} (\bibinfo {year} {2020})}\BibitemShut {NoStop}%
\bibitem [{\citenamefont {Rosenfeld}\ \emph {et~al.}(2021)\citenamefont {Rosenfeld}, \citenamefont {Riedinger}, \citenamefont {Gieseler}, \citenamefont {Schuetz},\ and\ \citenamefont {Lukin}}]{Rosenfeld2021-ef}%
  \BibitemOpen
  \bibfield  {author} {\bibinfo {author} {\bibfnamefont {E.}~\bibnamefont {Rosenfeld}}, \bibinfo {author} {\bibfnamefont {R.}~\bibnamefont {Riedinger}}, \bibinfo {author} {\bibfnamefont {J.}~\bibnamefont {Gieseler}}, \bibinfo {author} {\bibfnamefont {M.}~\bibnamefont {Schuetz}},\ and\ \bibinfo {author} {\bibfnamefont {M.~D.}\ \bibnamefont {Lukin}},\ }\bibfield  {title} {\bibinfo {title} {Efficient entanglement of spin qubits mediated by a hot mechanical oscillator},\ }\href {https://doi.org/10.1103/PhysRevLett.126.250505} {\bibfield  {journal} {\bibinfo  {journal} {Phys. Rev. Lett.}\ }\textbf {\bibinfo {volume} {126}},\ \bibinfo {pages} {250505} (\bibinfo {year} {2021})}\BibitemShut {NoStop}%
\bibitem [{\citenamefont {Bennett}\ \emph {et~al.}(2013{\natexlab{a}})\citenamefont {Bennett}, \citenamefont {Yao}, \citenamefont {Otterbach}, \citenamefont {Zoller}, \citenamefont {Rabl},\ and\ \citenamefont {Lukin}}]{Bennett2013PRL}%
  \BibitemOpen
  \bibfield  {author} {\bibinfo {author} {\bibfnamefont {S.~D.}\ \bibnamefont {Bennett}}, \bibinfo {author} {\bibfnamefont {N.~Y.}\ \bibnamefont {Yao}}, \bibinfo {author} {\bibfnamefont {J.}~\bibnamefont {Otterbach}}, \bibinfo {author} {\bibfnamefont {P.}~\bibnamefont {Zoller}}, \bibinfo {author} {\bibfnamefont {P.}~\bibnamefont {Rabl}},\ and\ \bibinfo {author} {\bibfnamefont {M.~D.}\ \bibnamefont {Lukin}},\ }\bibfield  {title} {\bibinfo {title} {Phonon-induced spin-spin interactions in diamond nanostructures: Application to spin squeezing},\ }\href {https://doi.org/10.1103/PhysRevLett.110.156402} {\bibfield  {journal} {\bibinfo  {journal} {Phys. Rev. Lett.}\ }\textbf {\bibinfo {volume} {110}},\ \bibinfo {pages} {156402} (\bibinfo {year} {2013}{\natexlab{a}})}\BibitemShut {NoStop}%
\bibitem [{\citenamefont {Barfuss}\ \emph {et~al.}(2015)\citenamefont {Barfuss}, \citenamefont {Teissier}, \citenamefont {Neu}, \citenamefont {Nunnenkamp},\ and\ \citenamefont {Maletinsky}}]{Barfuss2015NatPhys}%
  \BibitemOpen
  \bibfield  {author} {\bibinfo {author} {\bibfnamefont {A.}~\bibnamefont {Barfuss}}, \bibinfo {author} {\bibfnamefont {J.}~\bibnamefont {Teissier}}, \bibinfo {author} {\bibfnamefont {E.}~\bibnamefont {Neu}}, \bibinfo {author} {\bibfnamefont {A.}~\bibnamefont {Nunnenkamp}},\ and\ \bibinfo {author} {\bibfnamefont {P.}~\bibnamefont {Maletinsky}},\ }\bibfield  {title} {\bibinfo {title} {Strong mechanical driving of a single electron spin},\ }\href {https://doi.org/10.1038/nphys3411} {\bibfield  {journal} {\bibinfo  {journal} {Nat. Phys.}\ }\textbf {\bibinfo {volume} {11}},\ \bibinfo {pages} {820} (\bibinfo {year} {2015})}\BibitemShut {NoStop}%
\bibitem [{\citenamefont {Sasaki}\ and\ \citenamefont {Abe}(2024)}]{Sasaki2024PRL}%
  \BibitemOpen
  \bibfield  {author} {\bibinfo {author} {\bibfnamefont {K.}~\bibnamefont {Sasaki}}\ and\ \bibinfo {author} {\bibfnamefont {E.}~\bibnamefont {Abe}},\ }\bibfield  {title} {\bibinfo {title} {Suppression of pulsed dynamic nuclear polarization by many-body spin dynamics},\ }\href {https://doi.org/10.1103/PhysRevLett.132.106904} {\bibfield  {journal} {\bibinfo  {journal} {Phys. Rev. Lett.}\ }\textbf {\bibinfo {volume} {132}},\ \bibinfo {pages} {106904} (\bibinfo {year} {2024})}\BibitemShut {NoStop}%
\bibitem [{\citenamefont {Schwartz}\ \emph {et~al.}(2018)\citenamefont {Schwartz}, \citenamefont {Scheuer}, \citenamefont {Tratzmiller}, \citenamefont {Müller}, \citenamefont {Chen}, \citenamefont {Dhand}, \citenamefont {Wang}, \citenamefont {Müller}, \citenamefont {Naydenov}, \citenamefont {Jelezko},\ and\ \citenamefont {Plenio}}]{Schwartz2018}%
  \BibitemOpen
  \bibfield  {author} {\bibinfo {author} {\bibfnamefont {I.}~\bibnamefont {Schwartz}}, \bibinfo {author} {\bibfnamefont {J.}~\bibnamefont {Scheuer}}, \bibinfo {author} {\bibfnamefont {B.}~\bibnamefont {Tratzmiller}}, \bibinfo {author} {\bibfnamefont {S.}~\bibnamefont {Müller}}, \bibinfo {author} {\bibfnamefont {Q.}~\bibnamefont {Chen}}, \bibinfo {author} {\bibfnamefont {I.}~\bibnamefont {Dhand}}, \bibinfo {author} {\bibfnamefont {Z.-Y.}\ \bibnamefont {Wang}}, \bibinfo {author} {\bibfnamefont {C.}~\bibnamefont {Müller}}, \bibinfo {author} {\bibfnamefont {B.}~\bibnamefont {Naydenov}}, \bibinfo {author} {\bibfnamefont {F.}~\bibnamefont {Jelezko}},\ and\ \bibinfo {author} {\bibfnamefont {M.~B.}\ \bibnamefont {Plenio}},\ }\bibfield  {title} {\bibinfo {title} {Robust optical polarization of nuclear spin baths using hamiltonian engineering of nitrogen-vacancy center quantum dynamics},\ }\bibfield  {journal} {\bibinfo  {journal} {Sci. Adv.}\ }\textbf {\bibinfo {volume} {4}},\ \href
  {https://doi.org/10.1126/sciadv.aat8978} {10.1126/sciadv.aat8978} (\bibinfo {year} {2018})\BibitemShut {NoStop}%
\bibitem [{\citenamefont {Bartling}\ \emph {et~al.}(2023)\citenamefont {Bartling}, \citenamefont {Demetriou}, \citenamefont {Zutt}, \citenamefont {Kwiatkowski}, \citenamefont {Degen}, \citenamefont {Loenen}, \citenamefont {Bradley}, \citenamefont {Markham}, \citenamefont {Twitchen},\ and\ \citenamefont {Taminiau}}]{Bartling2023a}%
  \BibitemOpen
  \bibfield  {author} {\bibinfo {author} {\bibfnamefont {H.~P.}\ \bibnamefont {Bartling}}, \bibinfo {author} {\bibfnamefont {N.}~\bibnamefont {Demetriou}}, \bibinfo {author} {\bibfnamefont {N.~C.~F.}\ \bibnamefont {Zutt}}, \bibinfo {author} {\bibfnamefont {D.}~\bibnamefont {Kwiatkowski}}, \bibinfo {author} {\bibfnamefont {M.~J.}\ \bibnamefont {Degen}}, \bibinfo {author} {\bibfnamefont {S.~J.~H.}\ \bibnamefont {Loenen}}, \bibinfo {author} {\bibfnamefont {C.~E.}\ \bibnamefont {Bradley}}, \bibinfo {author} {\bibfnamefont {M.}~\bibnamefont {Markham}}, \bibinfo {author} {\bibfnamefont {D.~J.}\ \bibnamefont {Twitchen}},\ and\ \bibinfo {author} {\bibfnamefont {T.~H.}\ \bibnamefont {Taminiau}},\ }\bibfield  {title} {\bibinfo {title} {Control of individual electron-spin pairs in an electron-spin bath},\ }\href@noop {} {\bibfield  {journal} {\bibinfo  {journal} {arXiv: 2311.10110}\ } (\bibinfo {year} {2023})}\BibitemShut {NoStop}%
\bibitem [{\citenamefont {Babinec}\ \emph {et~al.}(2010)\citenamefont {Babinec}, \citenamefont {Hausmann}, \citenamefont {Khan}, \citenamefont {Zhang}, \citenamefont {Maze}, \citenamefont {Hemmer},\ and\ \citenamefont {Lončar}}]{loncar}%
  \BibitemOpen
  \bibfield  {author} {\bibinfo {author} {\bibfnamefont {T.~M.}\ \bibnamefont {Babinec}}, \bibinfo {author} {\bibfnamefont {B.~J.~M.}\ \bibnamefont {Hausmann}}, \bibinfo {author} {\bibfnamefont {M.}~\bibnamefont {Khan}}, \bibinfo {author} {\bibfnamefont {Y.}~\bibnamefont {Zhang}}, \bibinfo {author} {\bibfnamefont {J.~R.}\ \bibnamefont {Maze}}, \bibinfo {author} {\bibfnamefont {P.~R.}\ \bibnamefont {Hemmer}},\ and\ \bibinfo {author} {\bibfnamefont {M.}~\bibnamefont {Lončar}},\ }\bibfield  {title} {\bibinfo {title} {A diamond nanowire single-photon source},\ }\href {https://doi.org/10.1038/nnano.2010.6} {\bibfield  {journal} {\bibinfo  {journal} {Nat. Nanotechnol}\ }\textbf {\bibinfo {volume} {5}},\ \bibinfo {pages} {195–199} (\bibinfo {year} {2010})}\BibitemShut {NoStop}%
\bibitem [{\citenamefont {Wang}(2023)}]{Wang2023PhdThesis}%
  \BibitemOpen
  \bibfield  {author} {\bibinfo {author} {\bibfnamefont {Y.}~\bibnamefont {Wang}},\ }\emph {\bibinfo {title} {Using spins in diamond for quantum technologies}},\ \href {https://doi.org/10.4233/UUID:EC5051F5-0724-4921-8CBD-3C08B47328A6} {Ph.D. thesis},\ \bibinfo  {school} {Delft University of Technology} (\bibinfo {year} {2023})\BibitemShut {NoStop}%
\bibitem [{\citenamefont {Agarwal}(2010)}]{Agarwal2010-rf}%
  \BibitemOpen
  \bibfield  {author} {\bibinfo {author} {\bibfnamefont {G.~S.}\ \bibnamefont {Agarwal}},\ }\bibfield  {title} {\bibinfo {title} {Saving entanglement via a nonuniform sequence of $\pi$ pulses},\ }\href {https://doi.org/10.1088/0031-8949/82/03/038103} {\bibfield  {journal} {\bibinfo  {journal} {Phys. Scr.}\ }\textbf {\bibinfo {volume} {82}},\ \bibinfo {pages} {038103} (\bibinfo {year} {2010})}\BibitemShut {NoStop}%
\bibitem [{\citenamefont {Rao}\ \emph {et~al.}(2016)\citenamefont {Rao}, \citenamefont {Momenzadeh},\ and\ \citenamefont {Wrachtrup}}]{Rao2016-uw}%
  \BibitemOpen
  \bibfield  {author} {\bibinfo {author} {\bibfnamefont {D.~D.~B.}\ \bibnamefont {Rao}}, \bibinfo {author} {\bibfnamefont {S.~A.}\ \bibnamefont {Momenzadeh}},\ and\ \bibinfo {author} {\bibfnamefont {J.}~\bibnamefont {Wrachtrup}},\ }\bibfield  {title} {\bibinfo {title} {Heralded control of mechanical motion by single spins},\ }\href {https://doi.org/10.1103/PhysRevLett.117.077203} {\bibfield  {journal} {\bibinfo  {journal} {Phys. Rev. Lett.}\ }\textbf {\bibinfo {volume} {117}},\ \bibinfo {pages} {077203} (\bibinfo {year} {2016})}\BibitemShut {NoStop}%
\bibitem [{\citenamefont {Bhaktavatsala~Rao}\ \emph {et~al.}(2011)\citenamefont {Bhaktavatsala~Rao}, \citenamefont {Bar-Gill},\ and\ \citenamefont {Kurizki}}]{RaoPRL2011}%
  \BibitemOpen
  \bibfield  {author} {\bibinfo {author} {\bibfnamefont {D.~D.}\ \bibnamefont {Bhaktavatsala~Rao}}, \bibinfo {author} {\bibfnamefont {N.}~\bibnamefont {Bar-Gill}},\ and\ \bibinfo {author} {\bibfnamefont {G.}~\bibnamefont {Kurizki}},\ }\bibfield  {title} {\bibinfo {title} {Generation of macroscopic superpositions of quantum states by linear coupling to a bath},\ }\href {https://doi.org/10.1103/PhysRevLett.106.010404} {\bibfield  {journal} {\bibinfo  {journal} {Phys. Rev. Lett.}\ }\textbf {\bibinfo {volume} {106}},\ \bibinfo {pages} {010404} (\bibinfo {year} {2011})}\BibitemShut {NoStop}%
\bibitem [{\citenamefont {Taminiau}\ \emph {et~al.}(2014)\citenamefont {Taminiau}, \citenamefont {Cramer}, \citenamefont {van~der Sar}, \citenamefont {Dobrovitski},\ and\ \citenamefont {Hanson}}]{Taminiau2014}%
  \BibitemOpen
  \bibfield  {author} {\bibinfo {author} {\bibfnamefont {T.~H.}\ \bibnamefont {Taminiau}}, \bibinfo {author} {\bibfnamefont {J.}~\bibnamefont {Cramer}}, \bibinfo {author} {\bibfnamefont {T.}~\bibnamefont {van~der Sar}}, \bibinfo {author} {\bibfnamefont {V.~V.}\ \bibnamefont {Dobrovitski}},\ and\ \bibinfo {author} {\bibfnamefont {R.}~\bibnamefont {Hanson}},\ }\bibfield  {title} {\bibinfo {title} {Universal control and error correction in multi-qubit spin registers in diamond},\ }\href {https://doi.org/10.1038/nnano.2014.2} {\bibfield  {journal} {\bibinfo  {journal} {Nat. Nanotechnol.}\ }\textbf {\bibinfo {volume} {9}},\ \bibinfo {pages} {171} (\bibinfo {year} {2014})}\BibitemShut {NoStop}%
\bibitem [{\citenamefont {Bradley}\ \emph {et~al.}(2019)\citenamefont {Bradley}, \citenamefont {Randall}, \citenamefont {Abobeih}, \citenamefont {Berrevoets}, \citenamefont {Degen}, \citenamefont {Bakker}, \citenamefont {Markham}, \citenamefont {Twitchen},\ and\ \citenamefont {Taminiau}}]{Bradley2019}%
  \BibitemOpen
  \bibfield  {author} {\bibinfo {author} {\bibfnamefont {C.~E.}\ \bibnamefont {Bradley}}, \bibinfo {author} {\bibfnamefont {J.}~\bibnamefont {Randall}}, \bibinfo {author} {\bibfnamefont {M.~H.}\ \bibnamefont {Abobeih}}, \bibinfo {author} {\bibfnamefont {R.~C.}\ \bibnamefont {Berrevoets}}, \bibinfo {author} {\bibfnamefont {M.~J.}\ \bibnamefont {Degen}}, \bibinfo {author} {\bibfnamefont {M.~A.}\ \bibnamefont {Bakker}}, \bibinfo {author} {\bibfnamefont {M.}~\bibnamefont {Markham}}, \bibinfo {author} {\bibfnamefont {D.~J.}\ \bibnamefont {Twitchen}},\ and\ \bibinfo {author} {\bibfnamefont {T.~H.}\ \bibnamefont {Taminiau}},\ }\bibfield  {title} {\bibinfo {title} {A ten-qubit solid-state spin register with quantum memory up to one minute},\ }\href {https://doi.org/10.1103/PhysRevX.9.031045} {\bibfield  {journal} {\bibinfo  {journal} {Phys. Rev. X}\ }\textbf {\bibinfo {volume} {9}},\ \bibinfo {pages} {031045} (\bibinfo {year} {2019})}\BibitemShut {NoStop}%
\bibitem [{\citenamefont {Uhrig}(2007)}]{Uhrig2007-nu}%
  \BibitemOpen
  \bibfield  {author} {\bibinfo {author} {\bibfnamefont {G.~S.}\ \bibnamefont {Uhrig}},\ }\bibfield  {title} {\bibinfo {title} {Keeping a quantum bit alive by optimized pi-pulse sequences},\ }\href {https://doi.org/10.1103/PhysRevLett.98.100504} {\bibfield  {journal} {\bibinfo  {journal} {Phys. Rev. Lett.}\ }\textbf {\bibinfo {volume} {98}},\ \bibinfo {pages} {100504} (\bibinfo {year} {2007})}\BibitemShut {NoStop}%
\bibitem [{\citenamefont {Verbridge}\ \emph {et~al.}(2008)\citenamefont {Verbridge}, \citenamefont {Craighead},\ and\ \citenamefont {Parpia}}]{Verbridge2008}%
  \BibitemOpen
  \bibfield  {author} {\bibinfo {author} {\bibfnamefont {S.~S.}\ \bibnamefont {Verbridge}}, \bibinfo {author} {\bibfnamefont {H.~G.}\ \bibnamefont {Craighead}},\ and\ \bibinfo {author} {\bibfnamefont {J.~M.}\ \bibnamefont {Parpia}},\ }\bibfield  {title} {\bibinfo {title} {A megahertz nanomechanical resonator with room temperature quality factor over a million},\ }\bibfield  {journal} {\bibinfo  {journal} {Appl. Phys. Lett.}\ }\textbf {\bibinfo {volume} {92}},\ \href {https://doi.org/10.1063/1.2822406} {10.1063/1.2822406} (\bibinfo {year} {2008})\BibitemShut {NoStop}%
\bibitem [{\citenamefont {Li}\ \emph {et~al.}(2007)\citenamefont {Li}, \citenamefont {Tang},\ and\ \citenamefont {Roukes}}]{Li2007}%
  \BibitemOpen
  \bibfield  {author} {\bibinfo {author} {\bibfnamefont {M.}~\bibnamefont {Li}}, \bibinfo {author} {\bibfnamefont {H.~X.}\ \bibnamefont {Tang}},\ and\ \bibinfo {author} {\bibfnamefont {M.~L.}\ \bibnamefont {Roukes}},\ }\bibfield  {title} {\bibinfo {title} {Ultra-sensitive nems-based cantilevers for sensing, scanned probe and very high-frequency applications},\ }\href {https://doi.org/10.1038/nnano.2006.208} {\bibfield  {journal} {\bibinfo  {journal} {Nat. Nanotechnol.}\ }\textbf {\bibinfo {volume} {2}},\ \bibinfo {pages} {114} (\bibinfo {year} {2007})}\BibitemShut {NoStop}%
\bibitem [{\citenamefont {Gottesman}\ \emph {et~al.}(2001)\citenamefont {Gottesman}, \citenamefont {Kitaev},\ and\ \citenamefont {Preskill}}]{Gottesman2001-tl}%
  \BibitemOpen
  \bibfield  {author} {\bibinfo {author} {\bibfnamefont {D.}~\bibnamefont {Gottesman}}, \bibinfo {author} {\bibfnamefont {A.}~\bibnamefont {Kitaev}},\ and\ \bibinfo {author} {\bibfnamefont {J.}~\bibnamefont {Preskill}},\ }\bibfield  {title} {\bibinfo {title} {Encoding a qubit in an oscillator},\ }\href {https://doi.org/10.1103/PhysRevA.64.012310} {\bibfield  {journal} {\bibinfo  {journal} {Phys. Rev. A}\ }\textbf {\bibinfo {volume} {64}},\ \bibinfo {pages} {012310} (\bibinfo {year} {2001})}\BibitemShut {NoStop}%
\bibitem [{\citenamefont {Terhal}\ and\ \citenamefont {Weigand}(2016)}]{Terhal2016-sy}%
  \BibitemOpen
  \bibfield  {author} {\bibinfo {author} {\bibfnamefont {B.~M.}\ \bibnamefont {Terhal}}\ and\ \bibinfo {author} {\bibfnamefont {D.}~\bibnamefont {Weigand}},\ }\bibfield  {title} {\bibinfo {title} {Encoding a qubit into a cavity mode in circuit {QED} using phase estimation},\ }\href {https://doi.org/10.1103/PhysRevA.93.012315} {\bibfield  {journal} {\bibinfo  {journal} {Phys. Rev. A}\ }\textbf {\bibinfo {volume} {93}},\ \bibinfo {pages} {012315} (\bibinfo {year} {2016})}\BibitemShut {NoStop}%
\bibitem [{\citenamefont {Vuillot}\ \emph {et~al.}(2019)\citenamefont {Vuillot}, \citenamefont {Asasi}, \citenamefont {Wang}, \citenamefont {Pryadko},\ and\ \citenamefont {Terhal}}]{Vuillot2019_qec}%
  \BibitemOpen
  \bibfield  {author} {\bibinfo {author} {\bibfnamefont {C.}~\bibnamefont {Vuillot}}, \bibinfo {author} {\bibfnamefont {H.}~\bibnamefont {Asasi}}, \bibinfo {author} {\bibfnamefont {Y.}~\bibnamefont {Wang}}, \bibinfo {author} {\bibfnamefont {L.~P.}\ \bibnamefont {Pryadko}},\ and\ \bibinfo {author} {\bibfnamefont {B.~M.}\ \bibnamefont {Terhal}},\ }\bibfield  {title} {\bibinfo {title} {Quantum error correction with the toric gottesman-kitaev-preskill code},\ }\href {https://doi.org/10.1103/PhysRevA.99.032344} {\bibfield  {journal} {\bibinfo  {journal} {Phys. Rev. A}\ }\textbf {\bibinfo {volume} {99}},\ \bibinfo {pages} {032344} (\bibinfo {year} {2019})}\BibitemShut {NoStop}%
\bibitem [{\citenamefont {Duivenvoorden}\ \emph {et~al.}(2017)\citenamefont {Duivenvoorden}, \citenamefont {Terhal},\ and\ \citenamefont {Weigand}}]{Kasper2017_sensor}%
  \BibitemOpen
  \bibfield  {author} {\bibinfo {author} {\bibfnamefont {K.}~\bibnamefont {Duivenvoorden}}, \bibinfo {author} {\bibfnamefont {B.~M.}\ \bibnamefont {Terhal}},\ and\ \bibinfo {author} {\bibfnamefont {D.}~\bibnamefont {Weigand}},\ }\bibfield  {title} {\bibinfo {title} {Single-mode displacement sensor},\ }\href {https://doi.org/10.1103/PhysRevA.95.012305} {\bibfield  {journal} {\bibinfo  {journal} {Phys. Rev. A}\ }\textbf {\bibinfo {volume} {95}},\ \bibinfo {pages} {012305} (\bibinfo {year} {2017})}\BibitemShut {NoStop}%
\bibitem [{\citenamefont {Abobeih}\ \emph {et~al.}(2018)\citenamefont {Abobeih}, \citenamefont {Cramer}, \citenamefont {Bakker}, \citenamefont {Kalb}, \citenamefont {Markham}, \citenamefont {Twitchen},\ and\ \citenamefont {Taminiau}}]{Abobeih2018}%
  \BibitemOpen
  \bibfield  {author} {\bibinfo {author} {\bibfnamefont {M.~H.}\ \bibnamefont {Abobeih}}, \bibinfo {author} {\bibfnamefont {J.}~\bibnamefont {Cramer}}, \bibinfo {author} {\bibfnamefont {M.~A.}\ \bibnamefont {Bakker}}, \bibinfo {author} {\bibfnamefont {N.}~\bibnamefont {Kalb}}, \bibinfo {author} {\bibfnamefont {M.}~\bibnamefont {Markham}}, \bibinfo {author} {\bibfnamefont {D.~J.}\ \bibnamefont {Twitchen}},\ and\ \bibinfo {author} {\bibfnamefont {T.~H.}\ \bibnamefont {Taminiau}},\ }\bibfield  {title} {\bibinfo {title} {One-second coherence for a single electron spin coupled to a multi-qubit nuclear-spin environment},\ }\bibfield  {journal} {\bibinfo  {journal} {Nat. Commun.}\ }\textbf {\bibinfo {volume} {9}},\ \href {https://doi.org/10.1038/s41467-018-04916-z} {10.1038/s41467-018-04916-z} (\bibinfo {year} {2018})\BibitemShut {NoStop}%
\bibitem [{\citenamefont {Dasari}\ \emph {et~al.}(2022)\citenamefont {Dasari}, \citenamefont {Yang}, \citenamefont {Chakrabarti}, \citenamefont {Finkler}, \citenamefont {Kurizki},\ and\ \citenamefont {Wrachtrup}}]{durgancommn}%
  \BibitemOpen
  \bibfield  {author} {\bibinfo {author} {\bibfnamefont {D.~B.~R.}\ \bibnamefont {Dasari}}, \bibinfo {author} {\bibfnamefont {S.}~\bibnamefont {Yang}}, \bibinfo {author} {\bibfnamefont {A.}~\bibnamefont {Chakrabarti}}, \bibinfo {author} {\bibfnamefont {A.}~\bibnamefont {Finkler}}, \bibinfo {author} {\bibfnamefont {G.}~\bibnamefont {Kurizki}},\ and\ \bibinfo {author} {\bibfnamefont {J.}~\bibnamefont {Wrachtrup}},\ }\bibfield  {title} {\bibinfo {title} {Anti-zeno purification of spin baths by quantum probe measurements},\ }\bibfield  {journal} {\bibinfo  {journal} {Nat. Commun}\ }\textbf {\bibinfo {volume} {13}},\ \href {https://doi.org/10.1038/s41467-022-35045-3} {10.1038/s41467-022-35045-3} (\bibinfo {year} {2022})\BibitemShut {NoStop}%
\bibitem [{\citenamefont {Rao}\ \emph {et~al.}(2020)\citenamefont {Rao}, \citenamefont {Ghosh}, \citenamefont {Gelbwaser-Klimovsky}, \citenamefont {Bar-Gill},\ and\ \citenamefont {Kurizki}}]{Rao2020NJP}%
  \BibitemOpen
  \bibfield  {author} {\bibinfo {author} {\bibfnamefont {D.~D.~B.}\ \bibnamefont {Rao}}, \bibinfo {author} {\bibfnamefont {A.}~\bibnamefont {Ghosh}}, \bibinfo {author} {\bibfnamefont {D.}~\bibnamefont {Gelbwaser-Klimovsky}}, \bibinfo {author} {\bibfnamefont {N.}~\bibnamefont {Bar-Gill}},\ and\ \bibinfo {author} {\bibfnamefont {G.}~\bibnamefont {Kurizki}},\ }\bibfield  {title} {\bibinfo {title} {Spin-bath polarization via disentanglement},\ }\href {https://doi.org/10.1088/1367-2630/aba29a} {\bibfield  {journal} {\bibinfo  {journal} {New J. Phys.}\ }\textbf {\bibinfo {volume} {22}},\ \bibinfo {pages} {083035} (\bibinfo {year} {2020})}\BibitemShut {NoStop}%
\bibitem [{\citenamefont {Whiteley}\ \emph {et~al.}(2019)\citenamefont {Whiteley}, \citenamefont {Wolfowicz}, \citenamefont {Anderson}, \citenamefont {Bourassa}, \citenamefont {Ma}, \citenamefont {Ye}, \citenamefont {Koolstra}, \citenamefont {Satzinger}, \citenamefont {Holt}, \citenamefont {Heremans}, \citenamefont {Cleland}, \citenamefont {Schuster}, \citenamefont {Galli},\ and\ \citenamefont {Awschalom}}]{SiC}%
  \BibitemOpen
  \bibfield  {author} {\bibinfo {author} {\bibfnamefont {S.~J.}\ \bibnamefont {Whiteley}}, \bibinfo {author} {\bibfnamefont {G.}~\bibnamefont {Wolfowicz}}, \bibinfo {author} {\bibfnamefont {C.~P.}\ \bibnamefont {Anderson}}, \bibinfo {author} {\bibfnamefont {A.}~\bibnamefont {Bourassa}}, \bibinfo {author} {\bibfnamefont {H.}~\bibnamefont {Ma}}, \bibinfo {author} {\bibfnamefont {M.}~\bibnamefont {Ye}}, \bibinfo {author} {\bibfnamefont {G.}~\bibnamefont {Koolstra}}, \bibinfo {author} {\bibfnamefont {K.~J.}\ \bibnamefont {Satzinger}}, \bibinfo {author} {\bibfnamefont {M.~V.}\ \bibnamefont {Holt}}, \bibinfo {author} {\bibfnamefont {F.~J.}\ \bibnamefont {Heremans}}, \bibinfo {author} {\bibfnamefont {A.~N.}\ \bibnamefont {Cleland}}, \bibinfo {author} {\bibfnamefont {D.~I.}\ \bibnamefont {Schuster}}, \bibinfo {author} {\bibfnamefont {G.}~\bibnamefont {Galli}},\ and\ \bibinfo {author} {\bibfnamefont {D.~D.}\ \bibnamefont {Awschalom}},\ }\bibfield  {title} {\bibinfo {title} {Spin–phonon interactions in silicon
  carbide addressed by gaussian acoustics},\ }\href {https://doi.org/10.1038/s41567-019-0420-0} {\bibfield  {journal} {\bibinfo  {journal} {Nat. Phys.}\ }\textbf {\bibinfo {volume} {15}},\ \bibinfo {pages} {490–495} (\bibinfo {year} {2019})}\BibitemShut {NoStop}%
\bibitem [{\citenamefont {Robledo}\ \emph {et~al.}(2011)\citenamefont {Robledo}, \citenamefont {Childress}, \citenamefont {Bernien}, \citenamefont {Hensen}, \citenamefont {Alkemade},\ and\ \citenamefont {Hanson}}]{Robledo2011}%
  \BibitemOpen
  \bibfield  {author} {\bibinfo {author} {\bibfnamefont {L.}~\bibnamefont {Robledo}}, \bibinfo {author} {\bibfnamefont {L.}~\bibnamefont {Childress}}, \bibinfo {author} {\bibfnamefont {H.}~\bibnamefont {Bernien}}, \bibinfo {author} {\bibfnamefont {B.}~\bibnamefont {Hensen}}, \bibinfo {author} {\bibfnamefont {P.~F.~A.}\ \bibnamefont {Alkemade}},\ and\ \bibinfo {author} {\bibfnamefont {R.}~\bibnamefont {Hanson}},\ }\bibfield  {title} {\bibinfo {title} {High-fidelity projective read-out of a solid-state spin quantum register},\ }\href {https://doi.org/10.1038/nature10401} {\bibfield  {journal} {\bibinfo  {journal} {Nature}\ }\textbf {\bibinfo {volume} {477}},\ \bibinfo {pages} {574} (\bibinfo {year} {2011})}\BibitemShut {NoStop}%
\bibitem [{\citenamefont {Abobeih}\ \emph {et~al.}(2022)\citenamefont {Abobeih}, \citenamefont {Wang}, \citenamefont {Randall}, \citenamefont {Loenen}, \citenamefont {Bradley}, \citenamefont {Markham}, \citenamefont {Twitchen}, \citenamefont {Terhal},\ and\ \citenamefont {Taminiau}}]{Abobeih2022}%
  \BibitemOpen
  \bibfield  {author} {\bibinfo {author} {\bibfnamefont {M.~H.}\ \bibnamefont {Abobeih}}, \bibinfo {author} {\bibfnamefont {Y.}~\bibnamefont {Wang}}, \bibinfo {author} {\bibfnamefont {J.}~\bibnamefont {Randall}}, \bibinfo {author} {\bibfnamefont {S.~J.~H.}\ \bibnamefont {Loenen}}, \bibinfo {author} {\bibfnamefont {C.~E.}\ \bibnamefont {Bradley}}, \bibinfo {author} {\bibfnamefont {M.}~\bibnamefont {Markham}}, \bibinfo {author} {\bibfnamefont {D.~J.}\ \bibnamefont {Twitchen}}, \bibinfo {author} {\bibfnamefont {B.~M.}\ \bibnamefont {Terhal}},\ and\ \bibinfo {author} {\bibfnamefont {T.~H.}\ \bibnamefont {Taminiau}},\ }\bibfield  {title} {\bibinfo {title} {Fault-tolerant operation of a logical qubit in a diamond quantum processor},\ }\href {https://doi.org/10.1038/s41586-022-04819-6} {\bibfield  {journal} {\bibinfo  {journal} {Nature}\ }\textbf {\bibinfo {volume} {606}},\ \bibinfo {pages} {884} (\bibinfo {year} {2022})}\BibitemShut {NoStop}%
\bibitem [{\citenamefont {Aslam}\ \emph {et~al.}(2017)\citenamefont {Aslam}, \citenamefont {Pfender}, \citenamefont {Neumann}, \citenamefont {Reuter}, \citenamefont {Zappe}, \citenamefont {Fávaro~de Oliveira}, \citenamefont {Denisenko}, \citenamefont {Sumiya}, \citenamefont {Onoda}, \citenamefont {Isoya},\ and\ \citenamefont {Wrachtrup}}]{Aslam2017}%
  \BibitemOpen
  \bibfield  {author} {\bibinfo {author} {\bibfnamefont {N.}~\bibnamefont {Aslam}}, \bibinfo {author} {\bibfnamefont {M.}~\bibnamefont {Pfender}}, \bibinfo {author} {\bibfnamefont {P.}~\bibnamefont {Neumann}}, \bibinfo {author} {\bibfnamefont {R.}~\bibnamefont {Reuter}}, \bibinfo {author} {\bibfnamefont {A.}~\bibnamefont {Zappe}}, \bibinfo {author} {\bibfnamefont {F.}~\bibnamefont {Fávaro~de Oliveira}}, \bibinfo {author} {\bibfnamefont {A.}~\bibnamefont {Denisenko}}, \bibinfo {author} {\bibfnamefont {H.}~\bibnamefont {Sumiya}}, \bibinfo {author} {\bibfnamefont {S.}~\bibnamefont {Onoda}}, \bibinfo {author} {\bibfnamefont {J.}~\bibnamefont {Isoya}},\ and\ \bibinfo {author} {\bibfnamefont {J.}~\bibnamefont {Wrachtrup}},\ }\bibfield  {title} {\bibinfo {title} {Nanoscale nuclear magnetic resonance with chemical resolution},\ }\href {https://doi.org/10.1126/science.aam8697} {\bibfield  {journal} {\bibinfo  {journal} {Science}\ }\textbf {\bibinfo {volume} {357}},\ \bibinfo {pages} {67} (\bibinfo {year}
  {2017})}\BibitemShut {NoStop}%
\bibitem [{\citenamefont {Chikkaraddy}\ \emph {et~al.}(2017)\citenamefont {Chikkaraddy}, \citenamefont {Turek}, \citenamefont {Kongsuwan}, \citenamefont {Benz}, \citenamefont {Carnegie}, \citenamefont {van~de Goor}, \citenamefont {de~Nijs}, \citenamefont {Demetriadou}, \citenamefont {Hess}, \citenamefont {Keyser},\ and\ \citenamefont {Baumberg}}]{chikkaraddy2017}%
  \BibitemOpen
  \bibfield  {author} {\bibinfo {author} {\bibfnamefont {R.}~\bibnamefont {Chikkaraddy}}, \bibinfo {author} {\bibfnamefont {V.~A.}\ \bibnamefont {Turek}}, \bibinfo {author} {\bibfnamefont {N.}~\bibnamefont {Kongsuwan}}, \bibinfo {author} {\bibfnamefont {F.}~\bibnamefont {Benz}}, \bibinfo {author} {\bibfnamefont {C.}~\bibnamefont {Carnegie}}, \bibinfo {author} {\bibfnamefont {T.}~\bibnamefont {van~de Goor}}, \bibinfo {author} {\bibfnamefont {B.}~\bibnamefont {de~Nijs}}, \bibinfo {author} {\bibfnamefont {A.}~\bibnamefont {Demetriadou}}, \bibinfo {author} {\bibfnamefont {O.}~\bibnamefont {Hess}}, \bibinfo {author} {\bibfnamefont {U.~F.}\ \bibnamefont {Keyser}},\ and\ \bibinfo {author} {\bibfnamefont {J.~J.}\ \bibnamefont {Baumberg}},\ }\bibfield  {title} {\bibinfo {title} {Mapping nanoscale hotspots with single-molecule emitters assembled into plasmonic nanocavities using dna origami},\ }\href {https://doi.org/10.1021/acs.nanolett.7b04283} {\bibfield  {journal} {\bibinfo  {journal} {Nano Lett.}\ }\textbf
  {\bibinfo {volume} {18}},\ \bibinfo {pages} {405} (\bibinfo {year} {2017})}\BibitemShut {NoStop}%
\bibitem [{\citenamefont {Peng}\ \emph {et~al.}(2024)\citenamefont {Peng}, \citenamefont {Wu}, \citenamefont {Wang}, \citenamefont {Zhang}, \citenamefont {Geng}, \citenamefont {Dasari}, \citenamefont {Cleland},\ and\ \citenamefont {Wrachtrup}}]{Peng2024}%
  \BibitemOpen
  \bibfield  {author} {\bibinfo {author} {\bibfnamefont {R.}~\bibnamefont {Peng}}, \bibinfo {author} {\bibfnamefont {X.}~\bibnamefont {Wu}}, \bibinfo {author} {\bibfnamefont {Y.}~\bibnamefont {Wang}}, \bibinfo {author} {\bibfnamefont {J.}~\bibnamefont {Zhang}}, \bibinfo {author} {\bibfnamefont {J.}~\bibnamefont {Geng}}, \bibinfo {author} {\bibfnamefont {D.~B.~R.}\ \bibnamefont {Dasari}}, \bibinfo {author} {\bibfnamefont {A.~N.}\ \bibnamefont {Cleland}},\ and\ \bibinfo {author} {\bibfnamefont {J.}~\bibnamefont {Wrachtrup}},\ }\href {https://doi.org/10.48550/ARXIV.2409.12938} {\bibinfo {title} {Hybrid spin-phonon architecture for scalable solid-state quantum nodes}} (\bibinfo {year} {2024})\BibitemShut {NoStop}%
\bibitem [{\citenamefont {Bennett}\ \emph {et~al.}(2013{\natexlab{b}})\citenamefont {Bennett}, \citenamefont {Yao}, \citenamefont {Otterbach}, \citenamefont {Zoller}, \citenamefont {Rabl},\ and\ \citenamefont {Lukin}}]{lukinstrain}%
  \BibitemOpen
  \bibfield  {author} {\bibinfo {author} {\bibfnamefont {S.~D.}\ \bibnamefont {Bennett}}, \bibinfo {author} {\bibfnamefont {N.~Y.}\ \bibnamefont {Yao}}, \bibinfo {author} {\bibfnamefont {J.}~\bibnamefont {Otterbach}}, \bibinfo {author} {\bibfnamefont {P.}~\bibnamefont {Zoller}}, \bibinfo {author} {\bibfnamefont {P.}~\bibnamefont {Rabl}},\ and\ \bibinfo {author} {\bibfnamefont {M.~D.}\ \bibnamefont {Lukin}},\ }\bibfield  {title} {\bibinfo {title} {Phonon-induced spin-spin interactions in diamond nanostructures: Application to spin squeezing},\ }\href {https://doi.org/10.1103/PhysRevLett.110.156402} {\bibfield  {journal} {\bibinfo  {journal} {Phys. Rev. Lett.}\ }\textbf {\bibinfo {volume} {110}},\ \bibinfo {pages} {156402} (\bibinfo {year} {2013}{\natexlab{b}})}\BibitemShut {NoStop}%
\bibitem [{\citenamefont {Huillery}\ \emph {et~al.}(2020)\citenamefont {Huillery}, \citenamefont {Delord}, \citenamefont {Nicolas}, \citenamefont {Van Den~Bossche}, \citenamefont {Perdriat},\ and\ \citenamefont {H\'etet}}]{HuilleryPRB2020}%
  \BibitemOpen
  \bibfield  {author} {\bibinfo {author} {\bibfnamefont {P.}~\bibnamefont {Huillery}}, \bibinfo {author} {\bibfnamefont {T.}~\bibnamefont {Delord}}, \bibinfo {author} {\bibfnamefont {L.}~\bibnamefont {Nicolas}}, \bibinfo {author} {\bibfnamefont {M.}~\bibnamefont {Van Den~Bossche}}, \bibinfo {author} {\bibfnamefont {M.}~\bibnamefont {Perdriat}},\ and\ \bibinfo {author} {\bibfnamefont {G.}~\bibnamefont {H\'etet}},\ }\bibfield  {title} {\bibinfo {title} {Spin mechanics with levitating ferromagnetic particles},\ }\href {https://doi.org/10.1103/PhysRevB.101.134415} {\bibfield  {journal} {\bibinfo  {journal} {Phys. Rev. B}\ }\textbf {\bibinfo {volume} {101}},\ \bibinfo {pages} {134415} (\bibinfo {year} {2020})}\BibitemShut {NoStop}%
\bibitem [{\citenamefont {Schomburg}(2011)}]{Schomburg2011}%
  \BibitemOpen
  \bibfield  {author} {\bibinfo {author} {\bibfnamefont {W.~K.}\ \bibnamefont {Schomburg}},\ }\href {https://doi.org/10.1007/978-3-642-19489-4} {\emph {\bibinfo {title} {Introduction to Microsystem Design}}}\ (\bibinfo  {publisher} {Springer Berlin Heidelberg},\ \bibinfo {year} {2011})\BibitemShut {NoStop}%
\bibitem [{\citenamefont {Fung}\ \emph {et~al.}(2024)\citenamefont {Fung}, \citenamefont {Rosenfeld}, \citenamefont {Schaefer}, \citenamefont {Kabcenell}, \citenamefont {Gieseler}, \citenamefont {Zhou}, \citenamefont {Madhavan}, \citenamefont {Aslam}, \citenamefont {Yacoby},\ and\ \citenamefont {Lukin}}]{lukinprop}%
  \BibitemOpen
  \bibfield  {author} {\bibinfo {author} {\bibfnamefont {F.}~\bibnamefont {Fung}}, \bibinfo {author} {\bibfnamefont {E.}~\bibnamefont {Rosenfeld}}, \bibinfo {author} {\bibfnamefont {J.~D.}\ \bibnamefont {Schaefer}}, \bibinfo {author} {\bibfnamefont {A.}~\bibnamefont {Kabcenell}}, \bibinfo {author} {\bibfnamefont {J.}~\bibnamefont {Gieseler}}, \bibinfo {author} {\bibfnamefont {T.~X.}\ \bibnamefont {Zhou}}, \bibinfo {author} {\bibfnamefont {T.}~\bibnamefont {Madhavan}}, \bibinfo {author} {\bibfnamefont {N.}~\bibnamefont {Aslam}}, \bibinfo {author} {\bibfnamefont {A.}~\bibnamefont {Yacoby}},\ and\ \bibinfo {author} {\bibfnamefont {M.~D.}\ \bibnamefont {Lukin}},\ }\bibfield  {title} {\bibinfo {title} {Toward programmable quantum processors based on spin qubits with mechanically mediated interactions and transport},\ }\href {https://doi.org/10.1103/PhysRevLett.132.263602} {\bibfield  {journal} {\bibinfo  {journal} {Phys. Rev. Lett.}\ }\textbf {\bibinfo {volume} {132}},\ \bibinfo {pages} {263602} (\bibinfo {year}
  {2024})}\BibitemShut {NoStop}%
\bibitem [{\citenamefont {Gonzalez-Ballestero}\ \emph {et~al.}(2022)\citenamefont {Gonzalez-Ballestero}, \citenamefont {van~der Sar},\ and\ \citenamefont {Romero-Isart}}]{Gonzalez2022PRB}%
  \BibitemOpen
  \bibfield  {author} {\bibinfo {author} {\bibfnamefont {C.}~\bibnamefont {Gonzalez-Ballestero}}, \bibinfo {author} {\bibfnamefont {T.}~\bibnamefont {van~der Sar}},\ and\ \bibinfo {author} {\bibfnamefont {O.}~\bibnamefont {Romero-Isart}},\ }\bibfield  {title} {\bibinfo {title} {Towards a quantum interface between spin waves and paramagnetic spin baths},\ }\href {https://doi.org/10.1103/PhysRevB.105.075410} {\bibfield  {journal} {\bibinfo  {journal} {Phys. Rev. B}\ }\textbf {\bibinfo {volume} {105}},\ \bibinfo {pages} {075410} (\bibinfo {year} {2022})}\BibitemShut {NoStop}%
\bibitem [{\citenamefont {Rabl}\ \emph {et~al.}(2009)\citenamefont {Rabl}, \citenamefont {Cappellaro}, \citenamefont {Dutt}, \citenamefont {Jiang}, \citenamefont {Maze},\ and\ \citenamefont {Lukin}}]{Rabl2009}%
  \BibitemOpen
  \bibfield  {author} {\bibinfo {author} {\bibfnamefont {P.}~\bibnamefont {Rabl}}, \bibinfo {author} {\bibfnamefont {P.}~\bibnamefont {Cappellaro}}, \bibinfo {author} {\bibfnamefont {M.~V.~G.}\ \bibnamefont {Dutt}}, \bibinfo {author} {\bibfnamefont {L.}~\bibnamefont {Jiang}}, \bibinfo {author} {\bibfnamefont {J.~R.}\ \bibnamefont {Maze}},\ and\ \bibinfo {author} {\bibfnamefont {M.~D.}\ \bibnamefont {Lukin}},\ }\bibfield  {title} {\bibinfo {title} {Strong magnetic coupling between an electronic spin qubit and a mechanical resonator},\ }\href {https://doi.org/10.1103/physrevb.79.041302} {\bibfield  {journal} {\bibinfo  {journal} {Phys. Rev. B}\ }\textbf {\bibinfo {volume} {79}},\ \bibinfo {pages} {041302} (\bibinfo {year} {2009})}\BibitemShut {NoStop}%
\bibitem [{\citenamefont {Bild}\ \emph {et~al.}(2023)\citenamefont {Bild}, \citenamefont {Fadel}, \citenamefont {Yang}, \citenamefont {von L{\"u}pke}, \citenamefont {Martin}, \citenamefont {Bruno},\ and\ \citenamefont {Chu}}]{Bild2023-ta}%
  \BibitemOpen
  \bibfield  {author} {\bibinfo {author} {\bibfnamefont {M.}~\bibnamefont {Bild}}, \bibinfo {author} {\bibfnamefont {M.}~\bibnamefont {Fadel}}, \bibinfo {author} {\bibfnamefont {Y.}~\bibnamefont {Yang}}, \bibinfo {author} {\bibfnamefont {U.}~\bibnamefont {von L{\"u}pke}}, \bibinfo {author} {\bibfnamefont {P.}~\bibnamefont {Martin}}, \bibinfo {author} {\bibfnamefont {A.}~\bibnamefont {Bruno}},\ and\ \bibinfo {author} {\bibfnamefont {Y.}~\bibnamefont {Chu}},\ }\bibfield  {title} {\bibinfo {title} {Schr{\"o}dinger cat states of a 16-microgram mechanical oscillator},\ }\href {https://doi.org/10.1126/science.adf7553} {\bibfield  {journal} {\bibinfo  {journal} {Science}\ }\textbf {\bibinfo {volume} {380}},\ \bibinfo {pages} {274} (\bibinfo {year} {2023})}\BibitemShut {NoStop}%
\bibitem [{\citenamefont {Fl{\"u}hmann}\ \emph {et~al.}(2019)\citenamefont {Fl{\"u}hmann}, \citenamefont {Nguyen}, \citenamefont {Marinelli}, \citenamefont {Negnevitsky}, \citenamefont {Mehta},\ and\ \citenamefont {Home}}]{Fluhmann2019-jc}%
  \BibitemOpen
  \bibfield  {author} {\bibinfo {author} {\bibfnamefont {C.}~\bibnamefont {Fl{\"u}hmann}}, \bibinfo {author} {\bibfnamefont {T.~L.}\ \bibnamefont {Nguyen}}, \bibinfo {author} {\bibfnamefont {M.}~\bibnamefont {Marinelli}}, \bibinfo {author} {\bibfnamefont {V.}~\bibnamefont {Negnevitsky}}, \bibinfo {author} {\bibfnamefont {K.}~\bibnamefont {Mehta}},\ and\ \bibinfo {author} {\bibfnamefont {J.~P.}\ \bibnamefont {Home}},\ }\bibfield  {title} {\bibinfo {title} {Encoding a qubit in a trapped-ion mechanical oscillator},\ }\href {https://doi.org/10.1038/s41586-019-0960-6} {\bibfield  {journal} {\bibinfo  {journal} {Nature}\ }\textbf {\bibinfo {volume} {566}},\ \bibinfo {pages} {513} (\bibinfo {year} {2019})}\BibitemShut {NoStop}%
\bibitem [{\citenamefont {Wang}(2017)}]{Wang2017-im}%
  \BibitemOpen
  \bibfield  {author} {\bibinfo {author} {\bibfnamefont {Y.}~\bibnamefont {Wang}},\ }\emph {\bibinfo {title} {Quantum Error Correction with the {GKP} Code and Concatenation with Stabilizer Codes}},\ \href {http://arxiv.org/abs/1908.00147} {Master's thesis},\ \bibinfo  {school} {RWTH-AACHEN UNIVERSITY} (\bibinfo {year} {2017})\BibitemShut {NoStop}%
\bibitem [{\citenamefont {Sivak}\ \emph {et~al.}(2023)\citenamefont {Sivak}, \citenamefont {Eickbusch}, \citenamefont {Royer}, \citenamefont {Singh}, \citenamefont {Tsioutsios}, \citenamefont {Ganjam}, \citenamefont {Miano}, \citenamefont {Brock}, \citenamefont {Ding}, \citenamefont {Frunzio}, \citenamefont {Girvin}, \citenamefont {Schoelkopf},\ and\ \citenamefont {Devoret}}]{Sivak2023}%
  \BibitemOpen
  \bibfield  {author} {\bibinfo {author} {\bibfnamefont {V.~V.}\ \bibnamefont {Sivak}}, \bibinfo {author} {\bibfnamefont {A.}~\bibnamefont {Eickbusch}}, \bibinfo {author} {\bibfnamefont {B.}~\bibnamefont {Royer}}, \bibinfo {author} {\bibfnamefont {S.}~\bibnamefont {Singh}}, \bibinfo {author} {\bibfnamefont {I.}~\bibnamefont {Tsioutsios}}, \bibinfo {author} {\bibfnamefont {S.}~\bibnamefont {Ganjam}}, \bibinfo {author} {\bibfnamefont {A.}~\bibnamefont {Miano}}, \bibinfo {author} {\bibfnamefont {B.~L.}\ \bibnamefont {Brock}}, \bibinfo {author} {\bibfnamefont {A.~Z.}\ \bibnamefont {Ding}}, \bibinfo {author} {\bibfnamefont {L.}~\bibnamefont {Frunzio}}, \bibinfo {author} {\bibfnamefont {S.~M.}\ \bibnamefont {Girvin}}, \bibinfo {author} {\bibfnamefont {R.~J.}\ \bibnamefont {Schoelkopf}},\ and\ \bibinfo {author} {\bibfnamefont {M.~H.}\ \bibnamefont {Devoret}},\ }\bibfield  {title} {\bibinfo {title} {Real-time quantum error correction beyond break-even},\ }\href {https://doi.org/10.1038/s41586-023-05782-6} {\bibfield
  {journal} {\bibinfo  {journal} {Nature}\ }\textbf {\bibinfo {volume} {616}},\ \bibinfo {pages} {50} (\bibinfo {year} {2023})}\BibitemShut {NoStop}%
\bibitem [{\citenamefont {Tao}\ \emph {et~al.}(2014)\citenamefont {Tao}, \citenamefont {Boss}, \citenamefont {Moores},\ and\ \citenamefont {Degen}}]{degen}%
  \BibitemOpen
  \bibfield  {author} {\bibinfo {author} {\bibfnamefont {Y.}~\bibnamefont {Tao}}, \bibinfo {author} {\bibfnamefont {J.~M.}\ \bibnamefont {Boss}}, \bibinfo {author} {\bibfnamefont {B.~A.}\ \bibnamefont {Moores}},\ and\ \bibinfo {author} {\bibfnamefont {C.~L.}\ \bibnamefont {Degen}},\ }\bibfield  {title} {\bibinfo {title} {Single-crystal diamond nanomechanical resonators with quality factors exceeding one million},\ }\bibfield  {journal} {\bibinfo  {journal} {Nat. Commun}\ }\textbf {\bibinfo {volume} {5}},\ \href {https://doi.org/10.1038/ncomms4638} {10.1038/ncomms4638} (\bibinfo {year} {2014})\BibitemShut {NoStop}%
\end{thebibliography}%

\end{document}